\documentclass[11pt]{article}

\usepackage{amsmath}
\usepackage{amssymb}
\usepackage{amsthm}
\usepackage{mathtools,nccmath}

\usepackage{graphicx}

\usepackage{ulem} 

\usepackage[colorlinks,citecolor=blue,urlcolor=blue]{hyperref}

\usepackage{fullpage}

\usepackage{algorithm}
\usepackage{algpseudocode}  
\algrenewcommand\algorithmicrequire{\textbf{Input:}}
\algrenewcommand\algorithmicensure{\textbf{Output:}}
\newcommand{\Input}{\Require}
\newcommand{\Output}{\Ensure}

\usepackage{natbib}

\def\R{\mathbb{R}} 

\def\bc{\boldsymbol{c}}

\def\bh{\boldsymbol{h}}

\def\bq{\boldsymbol{q}}

\def\bs{\boldsymbol{s}}
\def\bt{\boldsymbol{t}}
\def\bu{\boldsymbol{u}}
\def\bv{\boldsymbol{v}}
\def\bw{\boldsymbol{w}}
\def\bx{\boldsymbol{x}}

\usepackage{bm}

\DeclareMathOperator*{\argmin}{argmin}

\usepackage[capitalize,noabbrev]{cleveref}

\usepackage{booktabs}
\usepackage{multirow}
\usepackage{threeparttable} 

\usepackage{subcaption}

\usepackage[dvipsnames]{xcolor}

\newcommand{\Pcal}{\mathcal{P}}

\theoremstyle{definition}

\newtheorem{remark}{Remark}

\title{Beyond fixed thresholds: optimizing summaries of wearable device data via piecewise linearization of quantile functions}
\author{Junyoung Park, Neo Kok, Irina Gaynanova  \thanks{Corresponding author, irinagn@umich.edu}}

\date{\normalsize Department of Biostatistics, University of Michigan, Ann Arbor, Michigan, USA}

\begin{document}

\maketitle

\begin{abstract}
Wearable devices, such as actigraphy monitors and continuous glucose monitors (CGMs), capture high-frequency data, which are often summarized by the percentages of time spent within fixed thresholds. For example, actigraphy data are categorized into sedentary, light, and moderate-to-vigorous activity, while CGM data are divided into hypoglycemia, normoglycemia, and hyperglycemia based on a standard glucose range of 70–180 mg/dL. Although scientific and clinical guidelines inform the choice of thresholds, it remains unclear whether this choice is optimal and whether the same thresholds should be applied across different populations. In this work, we define threshold optimality with loss functions that quantify discrepancies between the full empirical distributions of wearable device measurements and their discretizations based on specific thresholds. We introduce two loss functions: one that aims to accurately reconstruct the original distributions and another that preserves the pairwise sample distances. Using the Wasserstein distance as the base measure, we reformulate the loss minimization as optimal piecewise linearization of quantile functions. We solve this optimization via stepwise algorithms and differential evolution. We also formulate semi-supervised approaches where some thresholds are predefined based on scientific rationale. Applications to CGM datasets from diverse populations, including individuals with type 1 diabetes, type 2 diabetes, and normal glycemic control, demonstrate that data-driven thresholds vary by population, improve discriminative power, and yield stronger associations with clinical variables over fixed thresholds.

\end{abstract}


\section{Introduction}\label{sec: intro}

Advances in wearable device technology have opened up new avenues for healthcare research by enabling real-time monitoring of physiological variables outside traditional clinical settings. Devices such as actigraphy monitors and continuous glucose monitors (CGMs) generate high-frequency, individualized time series data that promise to offer new insights into a wearer’s health status. Actigraphy monitors capture detailed physical activity patterns \citep{fishmanAssociationObjectivelyMeasured2016, matabuenaDistributionalDataAnalysis2023}, while CGMs provide critical data for diabetes management by recording interstitial glucose levels at frequent time intervals \citep{brollInterpretingBloodGLUcose2021, battelinoContinuousGlucoseMonitoring2023}. With continued improvements in convenience and cost-effectiveness, wearable devices are being rapidly adopted in both clinical and personal health monitoring. Consequently, their data have significant potential for enhancing our understanding of human health and disease.

Despite this promise, wearable device data pose substantial analytical challenges due to their high measurement frequency and varying monitoring durations. Thus, researchers {often use fixed thresholds to summarize univariate wearable measurements.} Actigraphy data, which record movement intensities in activity count per minute, are often summarized into time spent in sedentary, light, moderate, and vigorous activity categories defined, for example, by thresholds at 100, {2020}, and 5999 activity counts \citep{troianoPhysicalActivityUnited2008, fishmanAssociationObjectivelyMeasured2016}. In diabetes studies, CGM data are similarly summarized via Time-in-Range (TIR) metrics, which indicate the percentage of time glucose levels fall within ranges defined by established consensus thresholds: less than 54 mg/dL, 54--69 mg/dL, 70--180 mg/dL, 181--250 mg/dL, and above 250 mg/dL \citep{danneInternationalConsensusUse2017, battelinoClinicalTargetsContinuous2019, battelinoContinuousGlucoseMonitoring2023}. {While thresholds may be derived differently depending on the domain---actigraphy thresholds may be specific to device and physiological contexts, whereas CGM thresholds are fixed by consensus---the resulting summaries are routinely used in clinical guidelines and downstream statistical analyses.}

{While threshold-based summaries are widely adopted, in practice, thresholds established in one context are fixed and routinely applied to different populations and study settings.
For example, the activity thresholds derived by \citet{chandlerClassificationPhysicalActivity2016} for 8--12-year-old children are fixed and reused in studies of younger children and older adolescents \citep{chenUnderstandingPhysicalActivity2020, islamAccelerometerMeasuredPhysicalActivity2024}.} In the context of CGM data, 
{while clinical targets (the desired percentage of time spent within a specific threshold range) are often adjusted based on patient risk, the thresholds defining those ranges remain almost entirely fixed across clinical practice and research in type 1 and type 2 diabetes \citep{wrightTimeRangeHow2020}, and even in populations without diabetes \citep{keshetCGMapCharacterizingContinuous2023}, although they were originally established for type 1 diabetes population \citep{agiostratidouStandardizingClinicallyMeaningful2017}.} Furthermore, \citet{kattaInterpretableCausalInference2024} highlighted that adjusting CGM thresholds, such as altering the ``healthy'' glucose range from 70--180 mg/dL to 70--140 mg/dL, can lead to substantially different interpretations of treatment effects, underscoring the importance of selecting the ``right'' thresholds {for a specific population and context.}

An emerging alternative to fixed threshold-based summaries is representing wearable device data as entire \textit{{univariate} distributions}. Distributional representations continuously extend threshold-based summaries by encoding the time spent within each infinitesimal range across the entire domain. This approach maintains greater data granularity than the conventional summaries, resulting in improved analytic performance \citep{matabuenaGlucodensitiesNewRepresentation2021, kattaInterpretableCausalInference2024}. However, distributional representations may be challenging to interpret and communicate in clinical settings, {whereas threshold-based metrics provide simple, interpretable summaries that facilitate clinical communications, such as guidelines for at least 150 minutes per week of moderate-to-vigorous physical activity or at least 70\% of time in the 70--180 mg/dL range \citep{fishmanAssociationObjectivelyMeasured2016,battelinoContinuousGlucoseMonitoring2023}.}

\begin{figure}[t!]
    \captionsetup[subfigure]{justification=centering}
    
    \centering
    \begin{minipage}[t]{0.325\linewidth}
        \vspace{1mm}
        \subfloat[][]{\includegraphics[width=\linewidth]{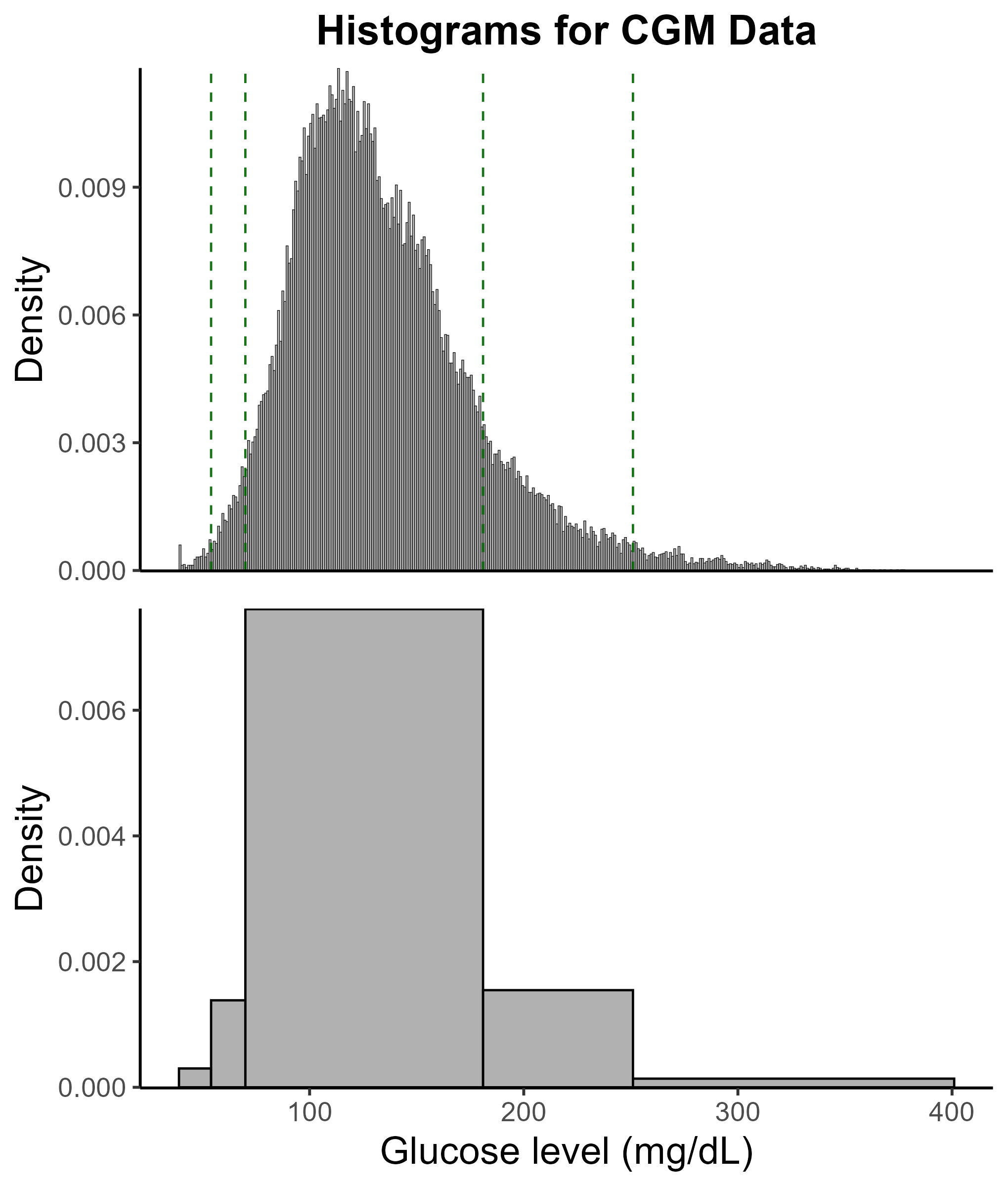}\label{fig: 1(a)}}
    \end{minipage}%
    \hfill
    \begin{minipage}[t]{0.65\linewidth}
        \vspace{-1mm}
        \subfloat[][]{\includegraphics[width=\linewidth]{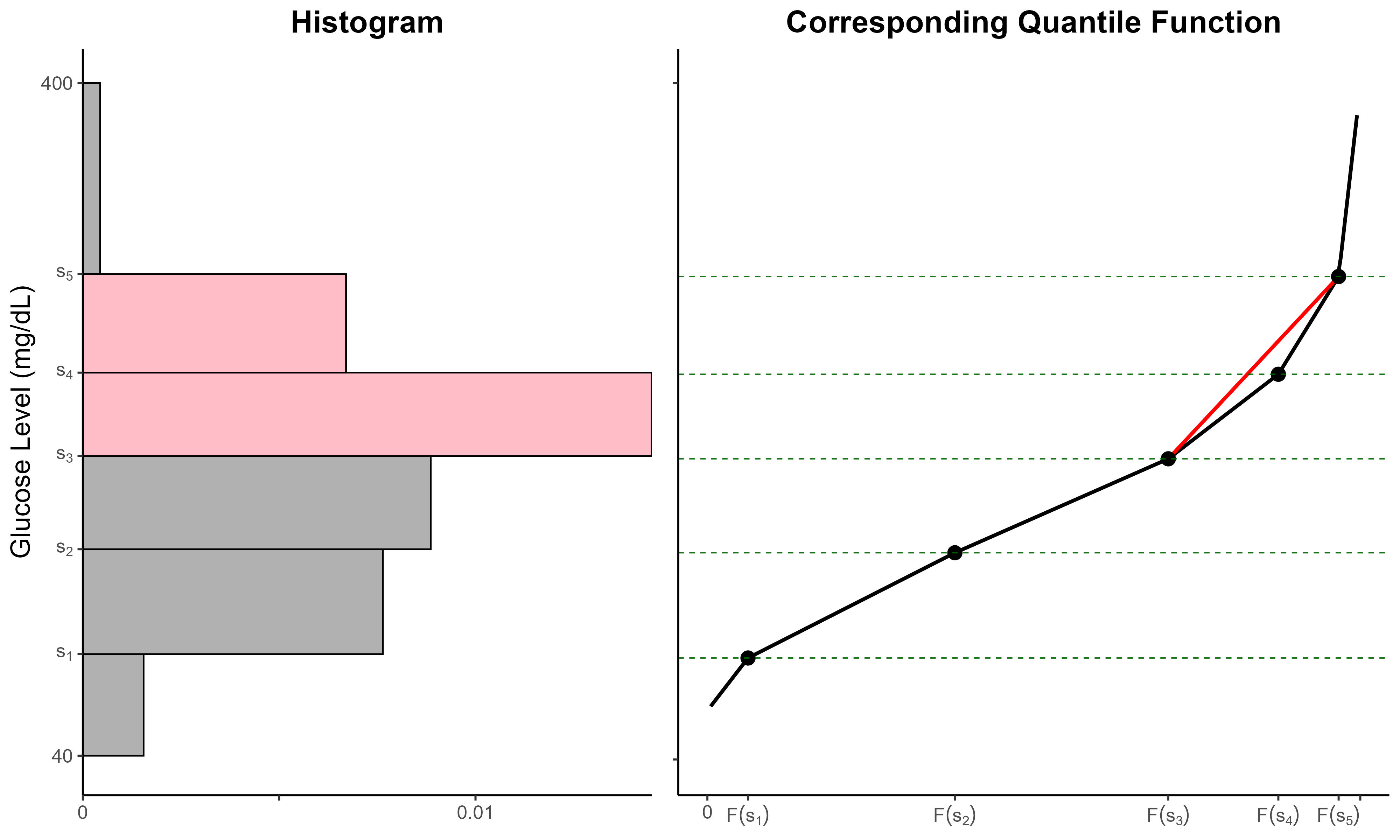}\label{fig: 1(b)}}
    \end{minipage}
    \caption{(a) Histograms of CGM measurements from a type 1 diabetes patient from \citet{brownSixMonthRandomizedMulticenter2019}. The top panel shows the full empirical distribution using a fixed bin width of 1 mg/dL, while the bottom panel amalgamates intermediate bins of four consensus thresholds---54, 70, 181, and 251 mg/dL---to create a coarser summary. 
    (b) Illustration of the quantile function (right) of a histogram (left) and their amalgamations. The amalgamation of red bins corresponds to the coarser linearization of the quantile function, depicted by the red line.}
    \label{fig: histogram-quantile}
\end{figure}

To balance the interpretability of fixed threshold-based summaries and the precision of distributional representations, we propose a framework for deriving data-dependent optimal thresholds. {We intentionally focus on unsupervised formulation as our primary motivating application is establishing thresholds for CGMs, where there is a known lack of studies with linked long-term outcomes such as mortality or diabetes-related complications \citep{gaynanovaDigitalBiomarkersGlucose2022}. Furthermore, like existing fixed thresholds, unsupervised data-driven thresholds remain consistent across outcomes and can be used within the same downstream statistical pipelines without compromising inferential validity.} Inspired by the concept of \textit{amalgamation} from compositional data analysis \citep{aitchisonStatisticalAnalysisCompositional1982, liPrincipalAmalgamationAnalysis2022}, which reduces dimension by merging compositional variables, we view threshold selection as amalgamating adjacent bins in a histogram of the full empirical distribution. Figure \ref{fig: 1(a)} illustrates this transformation of an empirical CGM distribution into a coarser histogram by amalgamating intermediate bins between selected thresholds. We then quantify threshold optimality using two loss functions: one measures the discrepancy between the original and amalgamated distributions, while the other preserves pairwise distances between distributions. 

Specifically, we use the Wasserstein distance as our primary discrepancy measure between {univariate} histograms, which quantifies the optimal transport cost between distributions \citep{villaniTopicsOptimalTransportation2003}. It effectively captures the geometry of the underlying domain of distributions and has been successfully applied to distributional data and histograms \citep{irpinoNewWassersteinBased2006, parkConvexClusteringAnalysis2019}. By leveraging the quantile-based computation of the Wasserstein distance for univariate distributions, we demonstrate a connection between histogram amalgamation and linear interpolation of quantiles (\cref{fig: 1(b)}), allowing us to reinterpret threshold selection as \textit{optimal piecewise linearization of quantile functions}. This perspective enables the efficient computation of the objective function of threshold optimization. 
{However, the underlying optimization remains a challenging discrete optimization over a large search space with constraints. As a practical baseline, we first propose greedy stepwise amalgamations, which are computationally feasible but may be suboptimal due to their reliance on local heuristics. To target global solutions, we also propose joint optimization via continuous relaxation; since this formulation remains non-convex and non-differentiable---rendering standard gradient-based methods ill-suited---we adopt the differential evolution algorithm \citep{dasDifferentialEvolutionSurvey2011}.}
Furthermore, our methods extend to a semi-supervised setting, which allows fixing certain clinically crucial thresholds while optimizing additional thresholds to better capture the distributional structures.

The proposed approach to data-driven thresholds fundamentally differs from the classical histogram binning methods \citep{freedmanHistogramDensityEstimator1981, knuthOptimalDatabasedBinning2019} and piecewise linearization \citep{nataliPiecewisePolynomialInterpolations2009}; these problems focus on a single distribution or function, whereas our work focuses on deriving optimal thresholds that apply to \textit{multiple distributions}. Additionally, while we adopt the concept of amalgamation from compositional data analysis, our approach goes beyond the purely compositional viewpoint often considered in actigraphy studies \citep{janssenSystematicReviewCompositional2020} by accounting for the inherent orderings of wearable device measurements. There are related recent works on data-driven thresholds by \citet{bankerSupervisedLearningPhysical2023} and \citet{bankerRegularizedScalaronfunctionRegression2024}, building on the supervised, scalar-on-function regression model with distributional predictors. However, {these thresholds are optimized for a specific continuous response, limiting their applicability in outcome-scarce settings or their generalizability across outcomes. Furthermore, they do not support valid post-selection inference on the same data without sample splitting, which complicates the analysis plan and reduces statistical power. In contrast, our unsupervised framework derives thresholds solely from the distributional characteristics of the data, ensuring they remain consistent across outcomes and can be used with standard inferential pipelines on the full dataset.}

In summary, our main contributions are: (1) a formulation of data-driven thresholds for optimal summaries of wearable device data, (2) a novel method for selecting thresholds with the piecewise-linear approximation of quantiles, and (3) two practical estimation approaches, stepwise algorithms and a global optimization with differential evolution, which can also accommodate semi-supervised scenarios. Analyses of real CGM data demonstrate that data-driven thresholds (i) differ considerably across populations with different diabetes status, (ii) improve discrimination in mixed populations, and (iii) capture stronger associations with lipid profiles---which are largely missed by consensus thresholds---among individuals without diabetes and those with early-stage type 2 diabetes. These results highlight the potential of our data-adaptive thresholds to improve statistical power in CGM-based analyses compared to fixed consensus thresholds.

\section{Optimal data-driven thresholds}\label{sec: optimal thresholds}

This section formalizes the concept of optimal data-driven thresholds {for summarizing univariate wearable measurements} based on the compositional structure of histogram data.

Our motivating example is CGM data, where glucose levels are recorded as integers from 40 to 400 mg/dL. The empirical distributional representation of these measurements for each individual is captured by a high-resolution histogram $h$, where each bin corresponds to a specific glucose value. Let ${h(j)}$ be the proportion of measurements at integer glucose level $j$, then histogram $h$ corresponds to a compositional vector: 
$$\bx = (h(40), h(41), \ldots,h(400))\in\Delta^{360},$$
where $\Delta^{p} = \{(z_0,\ldots,z_p) \in \R^{p+1}_{\ge 0}|\sum_{j=0}^pz_j = 1\}$ is the $p$-dimensional probability simplex. 

Using fixed consensus CGM thresholds of 54, 70, 181, and 251 mg/dL \citep{battelinoContinuousGlucoseMonitoring2023}, the corresponding summary of original histogram $h$ yields reduced compositional representation:
$$\left(\sum_{j=40}^{53} h(j), \sum_{j=54}^{69} h(j), \sum_{j=70}^{180} h(j), \sum_{j=181}^{250} h(j), \sum_{j=251}^{400} h(j)\right)\in\Delta^4,$$ where each component indicates Time-in-Range (TIR) proportions for 40--53, 54--69, 70--180, 181--250, and 251--400 mg/dL, respectively. This process aligns with amalgamation in compositional data analysis \citep{aitchisonStatisticalAnalysisCompositional1982}, with the additional constraint of merging only neighboring components to reflect the natural ordering of glucose values (\cref{fig: 1(a)}). Thus, a threshold-based summary inherently performs dimension reduction on the compositional level, so we can formulate the optimal threshold problem as a dimension reduction problem: reducing the number of bins while aiming to preserve the key characteristics of the original histogram $h$.

To generalize this idea, let $h_1,\ldots, h_n$ denote histogram data on a bounded interval $\Omega = [a, b]$ with bins separated by a sequence of cutoffs
$$a = s_0 < s_1 < \cdots < s_J < s_{J+1} =b.$$
{We use the subscript $i$ to index subjects throughout the paper.}
{Viewing each histogram $h_i$ as a density function constant over intervals $[s_j, s_{j+1})$, we let $x_{ij} = (s_{j+1} - s_j)h_i(s_j)$ denote the TIR proportion within $[s_j, s_{j+1})$ for the $i$th subject ($j=0,\ldots, J$), yielding
the compositional vector $\bx_i = (x_{i0}, \ldots, x_{iJ})\in\Delta^{J}$ that represents $h_i$}. 
Then, each \textit{subsequence} $\bt = \{t_1,\ldots,t_K\}$ of the cutoffs $\bs = \{s_1, \ldots, s_J\}$, where $s_{c_k} = t_k$ with $0 = c_0 < c_1< \cdots < c_{K} < c_{K+1} = J+1$, forms amalgamations:
$$\bx_i^{(\bt)} = \left(\sum_{j=c_0+1}^{c_1}x_{ij}, \sum_{j=c_1+1}^{c_2}x_{ij},\ldots,\sum_{j=c_{K}+1}^{c_{K+1}}x_{ij}\right)\in\Delta^{K},$$
where each component $\sum_{j=c_k + 1}^{c_{k+1}} x_{ij}$ indicates the TIR proportion within $[t_k, t_{k+1})$, with $t_0 = a$ and $t_{K+1} = b$. We denote the amalgamated histogram corresponding to $\bx_i^{(\bt)}$ by $h_i^{(\bt)}$.

Then, we define two types of optimality measures for thresholds $\bt$ for data $\bh =\{h_1,\ldots, h_n\}$, using a generic discrepancy measure between histograms $d(\cdot, \cdot)$:
\begin{enumerate}
    \item Distribution preservation:
    \begin{equation}\label{eq:loss_l1}
        L_1(\bt;\bh) := \frac{1}{n}\sum_{i=1}^n d^2(h_i, h_i^{(\bt)})
    \end{equation}

    \item Distance preservation:
    \begin{equation}\label{eq:loss_l2}
        L_2(\bt; \bh) := \frac{2}{n(n-1)}\sum_{i<i'} \left\{d(h_i, h_{i'}) - d(h_i^{(\bt)}, h_{i'}^{(\bt)})\right\}^2
    \end{equation}
\end{enumerate}
These loss functions are motivated by the dimension reduction context, thus aiming to recover the original structure of data $\bh$. 
{The first loss, which minimizes the average distance between $h_i$ and $h_i^{(\bt)}$, prioritizes faithful reconstruction of the original individual distributions. It is therefore well-suited for deriving a parsimonious summary of a single cohort that shares similar characteristics (e.g., individuals with the same disease).}
The second loss, though computationally more intensive, aims to preserve the pairwise distances between original distributions $h_1,\ldots, h_n$.
{This objective, similarly designed to multidimensional scaling \citep{borgModernMultidimensionalScaling2007}, is well-suited for capturing distinct subgroups (e.g., individuals with different disease status) or continuous gradients (e.g., disease progression severity).}
Similar objectives have been used in the amalgamation of compositional data \citep{liPrincipalAmalgamationAnalysis2022}.

Using these loss functions, we formulate the problem of selecting $K$ optimal data-driven thresholds as the following optimization problem
\begin{equation}\label{eq: opt-discrete}
   \widehat \bt = \underset{\bt\subset\bs,\, |\bt| = K}{\argmin}\,L(\bt;\mathbf{h}),
\end{equation}
where $L$ is either $L_1$~\eqref{eq:loss_l1} or $L_2$~\eqref{eq:loss_l2}, and $\bt\subset\bs$ indicates that $\bt$ is a monotonic subsequence of the high-resolution cutoffs $\bs = \{s_1,\ldots, s_J\} $. While this formulation is primarily motivated by integer-valued wearable device data---such as CGM data, where empirical distributions inherently define integer cutoffs $\bs$---it extends to continuous data; see the continuous relaxation in \cref{sec: continuous-relaxation}.

\subsection{Wasserstein distance}\label{sec: wasserstein}

We propose to use the Wasserstein distance as the discrepancy measure $d(\cdot, \cdot)$ {between univariate histograms} in our loss functions, which is the optimal transport metric between probability distributions \citep{villaniTopicsOptimalTransportation2003}. Unlike compositional distances \citep{aitchisonCriteriaMeasuresCompositional1992} or the $L^2$ distance between densities, the Wasserstein distance captures the geometry of the underlying space of distribution functions and thus leverages the additional structural information encoded by histograms beyond simple compositions.

Let $\Pcal_2(\Omega)$ denote the set of probability measures on a bounded interval $\Omega=[a, b]$ with finite second moments. The 2-Wasserstein distance $d_W$ between two measures $\mu, \nu \in\Pcal_2(\Omega) $ is defined as:
$$d^2_W(\mu, \nu) = \inf_{\gamma\in\Pi(\mu,\nu)}\int_{\Omega\times\Omega}|x-y|^2d\gamma(x, y),$$
where $\Pi(\mu, \nu)$ is the set of joint distributions on $\Omega\times\Omega$ having $\mu$ and $\nu$ as marginals. The distance $d_W(\mu,\nu)$ represents the minimum ``cost'' of transporting mass from one distribution to another, where the cost of transporting a unit mass between points $x$ and $y$ is proportional to the squared Euclidean distance $|x-y|^2$ on $\Omega$ \citep{villaniTopicsOptimalTransportation2003}. {In general, replacing this cost with $|x-y|^p$ ($p\ge 1$) yields the general $p$-Wasserstein distance that can also be used in our proposed optimizations. We focus on the 2-Wasserstein distance since it is a common choice in metric-based distributional data analysis \citep{parkConvexClusteringAnalysis2019,petersenFrechetRegressionRandom2019, coulterFastVariableSelection2025}.}

For one-dimensional distributions, the computation of the Wasserstein distance simplifies significantly by leveraging quantile functions. For any probability measure $\mu\in\Pcal_2(\Omega)$, let $F_\mu$ denote the CDF and $q_\mu(p) = \inf\{x\in\Omega: F_\mu(x)\ge p\}$ denote the quantile function. 
Then the 2-Wasserstein distance between $\mu$ and $\nu$ is computed as:
\begin{equation}\label{eq: L2-quantile}
    d_W^2(\mu,\nu) = \int_0^1 (q_\mu(p) -q_\nu(p))^2dp.
\end{equation}

\subsection{Quantile piecewise linearization}\label{sec: quantile-linearization}

The quantile-based computation of {univariate} Wasserstein distance \eqref{eq: L2-quantile} offers a new perspective on the threshold selection problem within our optimization framework \eqref{eq: opt-discrete}. Specifically, we further show that identifying optimal thresholds can be reframed as finding knots that yield the best piecewise linear approximations of the empirical quantile functions. Notably, the same knots are applied across all empirical quantile functions in the sample.

\cref{fig: 1(b)} illustrates this concept with a histogram $h$ on $[40, 400]$ with six bins, separated by thresholds $s_1<\cdots < s_5$, alongside its corresponding quantile function $q$. The histogram is regarded as a locally constant density, having a piecewise linear quantile function shown in the right panel. In this example, we consider an amalgamation of the red adjacent bins thresholded by $s_3, s_4$, and $s_5$. This removes the intermediate threshold $s_4$ at the histogram level. At the quantile level, this operation corresponds to a \textit{linear interpolation} between the points $(F(s_3), s_3)$ and $(F(s_5), s_5)$, where $F$ denotes the corresponding cumulative distribution function (CDF). This interpolation is represented by the red line in the right panel.

This insight reinterprets the amalgamation-based threshold optimization framework as a problem of coarser \textit{piecewise linearization} of quantile functions. In general, let $h_1,\ldots,h_n$ be histograms with bins separated by thresholds $\bs=\{s_1,\ldots,s_J\}$ and let $\bt = \{t_1,\ldots,t_K\}$ be a subsequence of thresholds $\bs$ as defined in \cref{sec: optimal thresholds}. Denoting $F_i$ by the CDF of each $h_i$, the quantile function $q_i^{(\bt)}$ of the amalgamated summary $h_i^{(\bt)}$ is constructed as a linear interpolation of the points:
$$(0, q_i(0)), (F_i(t_1), q_i\circ F_i(t_1)), \ldots, (F_i(t_{K}), q_i\circ F_i(t_{K})), (1, q_i(1)),$$ 
{where $q_i\circ F_i (t_k) := q_i(F_i(t_k))$.}
Specifically, for any $p\in (0, 1)$, let $k$ be the index of the threshold $t_k$ for which $p\in [F_i(t_k), F_i(t_{k+1})]$, with $F_i(t_{k+1})\neq F_i(t_k)$. Then the piecewise linear quantile function $q_i^{(\bt)}(p)$ can be evaluated as:
\begin{equation}\label{eq: quantile-interpolation}
    q_i^{(\bt)}(p) = q_i\circ F_i(t_k)+ \frac{q_i\circ F_i(t_{k+1}) - q_i\circ F_i(t_{k})}{F_i(t_{k+1}) - F_i(t_k)}(p - F_i(t_k)).
\end{equation}

Using these piecewise-interpolated quantile evaluations, the {2-}Wasserstein distance between {univariate} distributions can be computed via the $L^2$-distances between quantiles \eqref{eq: L2-quantile}. We approximate the integral by evaluating the quantile functions on a grid of $M$ equally spaced points within $(0, 1)$, denoted by $u_m = m/(M+1)$ for $m=1,\ldots, M$. A similar discretization approach for evaluating {univariate} Wasserstein distance has been considered in \citet{petersenFrechetRegressionRandom2019, coulterFastVariableSelection2025}. The quantile values on the grid form vectors $\bq_i = (q_i(u_1),\ldots,q_i(u_M))$ and $\bq_i^{(\bt)} = (q_i^{(\bt)}(u_1), \ldots, q_i^{(\bt)}(u_M))$ in $\R^{M}$, leading to the following approximation of our loss functions $L_1$~\eqref{eq:loss_l1} and $L_2$~\eqref{eq:loss_l2}:
\begin{align}
    L_1(\bt;\bh) &\approx \frac{1}{n}\sum_{i=1}^n\|\bq_i - \bq_i^{(\bt)}\|^2 / (M+1), \qquad \text{and} \label{eq: loss1} \\
    L_2(\bt;\bh) &\approx \frac{2}{n(n-1)} \sum_{i<i'} \left\{ \|\bq_i - \bq_{i'}\| - \|\bq_i^{(\bt)} - \bq_{i'}^{(\bt)}\| \right\}^2 / (M+1). \label{eq: loss2}
\end{align}
Thus, the optimal thresholds $\widehat \bt$ in~\eqref{eq: opt-discrete} correspond to the knots in piecewise linearizations of quantiles $q_1,\ldots, q_n$ that minimize the corresponding loss functions. In Section~\ref{sec: algorithm}, we describe algorithms to solve the optimization problem~\eqref{eq: opt-discrete}.

\begin{remark}
    {The proposed optimization~\eqref{eq: opt-discrete} requires fixing the number $K$ of thresholds a priori. Under our piecewise linearization formulation, larger $K$ always yields lower loss values by more closely approximating the full distributional signal, whereas smaller $K$ gives coarser but more interpretable summaries. Thus, analogous to principal component analysis, a practical data-driven strategy for choosing $K$ is to balance interpretability with the preservation of the underlying distributional structure. We propose to do this by inspecting the screeplot of the loss function as a function of $K$ and then applying the elbow rule to select the best $K$. Section~B.1 of the supplementary material provides examples of such screeplots for the real datasets analyzed in Section~\ref{sec: real-data}.}
\end{remark}

\subsection{Semi-supervised approach}\label{sec: semi-supervised}

In certain cases, domain knowledge or clinical guidelines may suggest fixing some specific thresholds to maintain clinically relevant information. For instance, in CGM data, thresholds at 70 and 181 mg/dL are often critical as many clinical targets of glucose control are based on the range of 70--180 mg/dL \citep{battelinoClinicalTargetsContinuous2019}. To accommodate these needs, we propose a \textit{semi-supervised approach} that optimizes the additional thresholds while retaining practically important thresholds. 

Following the notations of \eqref{eq: opt-discrete}, let $\bt_{\text{fix}} = \{t_1^*,\ldots, t_m^*\} \subset \bs$ denote predetermined thresholds set by practitioners. We aim to identify additional thresholds $\bt_{\text{opt}} = \{t_{m + 1},\ldots,t_K\}\subset\bs\setminus \bt_{\text{fix}}$ that refine the amalgamations of histograms $\bh$ by the combined thresholds $\bt = \bt_{\text{opt}} \cup \bt_{\text{fix}}$, which is sorted to compute piecewise linearizations $q_i^{(\bt)}$ \eqref{eq: quantile-interpolation}. Using the same quantile-based loss computations $L_1$ \eqref{eq: loss1} or $L_2$ \eqref{eq: loss2}, denoted by $L$, the semi-supervised optimization is formulated as:
\begin{equation}\label{eq: opt-semisuper}
    \widehat \bt = \underset{\bt_{\text{fix}}\subset\bt\subset\bs,\, |\bt| = K}{L(\bt;\bh)}
\end{equation}
Our proposed algorithms in \cref{sec: algorithm} accommodate the additional constraint of $\bt_{\text{fix}}\subset \bt$. We illustrate the semi-supervised analysis with real CGM data in \cref{sec: real-data}, where we fix $\bt_{\text{fix}}=\{70, 181\}$ mg/dL and identify additional optimal thresholds.

\section{Algorithm}\label{sec: algorithm}

In this section, we detail algorithms for solving the threshold selection problem \eqref{eq: opt-discrete} with the quantile-based computations of optimality criteria \eqref{eq: loss1} and \eqref{eq: loss2}. 
{Although the problem is mathematically well-defined, its combinatorial nature makes optimization computationally challenging. To solve this, we consider two approaches: greedy stepwise algorithms based on local heuristics, and a more principled joint optimization based on continuous relaxation.}

\subsection{Stepwise algorithms}\label{sec: stepwise algorithms}

The optimization \eqref{eq: opt-discrete} is a challenging combinatorial task. Given $J$ original cutoffs $a = s_0 < s_1 < \cdots < s_J < s_{J+1} =b$ and $K$ thresholds to optimize, the problem has discrete search space of size ${J}\choose{K}$, which becomes computationally infeasible for large $J$ and moderate $K$. For instance, the motivating example of CGM data with $J = 360$ and typical $K = 4$ gives over 688 million combinations. To maintain computational feasibility, we consider two iterative approaches based on merging or refining existing histogram bins: \textit{stepwise aggregation} and \textit{stepwise splitting}. These approaches are analogous to backward elimination and forward selection procedures for selecting the best subset of thresholds $\bs = \{s_1, \ldots, s_{J}\}$:
\begin{itemize}
    \item \textbf{Stepwise Aggregation}: 
    \begin{enumerate}
        \item Start with the full set of thresholds $\bt=\bs$.
        \item Remove the threshold $s_j\in\bt$ that minimizes $L(\bt\setminus \{s_j\};\bh)$, which aggregates neighboring bins of $\bh$ separated by $s_j$.
        \item Iterate step 2 until the number of thresholds $|\bt|$ reduces to $K$.
    \end{enumerate}
    \item \textbf{Stepwise Splitting}: 
    \begin{enumerate}
        \item Start with the empty set of thresholds $\bt=\emptyset$.
        \item Add the threshold $s_j\in\bs$ to $\bt$ that minimizes $L(\bt;\bh)$, which splits an existing bin of each $h_i$ into two parts.
        \item Iterate step 2 until the number of thresholds $|\bt|$ reaches $K$.
    \end{enumerate} 
\end{itemize}
With the quantile-based computations described in \cref{sec: quantile-linearization}, these stepwise algorithms can also be interpreted as iterative coarsening or refinement of piecewise linear approximations of quantile functions. Stepwise splitting is computationally more efficient for small $K$, whereas stepwise aggregation generally yields better solutions since thresholds are evaluated in the presence of all others; this behavior is similar to variable selection problems \citep{guyonIntroductionVariableFeature2003}. We compare the empirical performance of these two approaches in \cref{sec: simulations}.

\subsection{Joint optimization with continuous relaxation}\label{sec: continuous-relaxation}

Although stepwise algorithms are computationally feasible, they may produce suboptimal results due to their local, greedy nature. To address this limitation, we also propose a joint optimization approach based on a continuous relaxation of the threshold selection problem. While our focus is on the joint optimization of histograms from integer-valued measurements, such as our motivating CGM data, the continuously relaxed formulation inherently extends our threshold optimization methodology to continuous, non-integer distributions.  

In the quantile discretization formulation in \cref{sec: quantile-linearization}, we now allow the thresholds $\bt =\{t_1,\ldots,t_K\}$ to vary continuously across the range $\Omega=[a, b]$, subject to the monotonicity constraint $a < t_1 < \cdots < t_K < b$, instead of restricting them to a subset of the original discrete cutoffs $\bs=\{s_1,\ldots, s_J\}$. The linearly interpolated quantiles $q_i^{(\bt)}$ are computed as in \eqref{eq: quantile-interpolation}, formulating the continuous optimization as:
\begin{equation}\label{eq: opt-continuous}
    \widehat\bt = \underset{\bt\subset\Omega,\, |\bt| = K}{\argmin}\,L(\bt;\bh),
\end{equation}
subject to the constraint $a < t_1 < \cdots < t_K < b$, where $L$ is one of the loss functions \eqref{eq: loss1} or \eqref{eq: loss2}. The semi-supervised approach \eqref{eq: opt-semisuper} similarly extends to continuous fixed thresholds $\bt_{\text{fix}}\subset\Omega$, where the optimization follows \eqref{eq: opt-continuous} with the additional constraint $\bt_{\text{fix}} \subset \bt$. At the histogram level, the summarized histograms $h_i^{(\bt)}$ with continuous thresholds $\bt$ have compositional expressions:
\begin{equation*}
    \bx_i^{(\bt)}=\left(\int_{t_0}^{t_1}h_i(z)dz, \int_{t_1}^{t_2}h_i(z)dz, \ldots, \int_{t_{K}}^{t_{K+1}}h_i(z)dz \right) \in\Delta^{K},
\end{equation*}
where $h_i$ are regarded as locally constant densities, $t_0=a$ and $t_{K+1}=b$. This representation generalizes amalgamation into a \textit{soft amalgamation}, where the thresholds $\bt$ subdivide the original bins divided by $\bs$.

The new continuously relaxed problem \eqref{eq: opt-continuous} is still challenging from the optimization viewpoint. 
{First, it is nonconvex because $q_i^{(\mathbf{t})}$ depends on the CDFs $F_i$ through the denominator in (5). Second, it is not differentiable since the $F_i$ are non-differentiable at the histogram knots $s_1,\ldots, s_J$. Third, the thresholds are constrained to be in ascending order. These features make standard optimization methods, such as gradient-based approaches, ill-suited.}

To overcome these challenges, we propose to adopt \textit{differential evolution} (DE), 
{a principled evolutionary algorithm for global optimization. By stochastically refining a set of candidate solutions using only loss evaluations, DE avoids the need for gradient computations. Furthermore, while convergence to global optima is generally not guaranteed for standard methods in nonconvex optimization,} 
DE has been proven effective for problems with a small number of optimization variables, where it efficiently finds global or near-global optima on nonconvex landscapes while adhering to problem constraints \citep{dasDifferentialEvolutionSurvey2011}. 
{This makes it well-suited for our wearable device context: the number of thresholds $K$ is typically small for interpretability, commonly $K=2, 4$ for CGM data and $K=3$ for actigraphy summaries. A related recent work by \citet{brubakerFrequencyBandAnalysis2026} utilizes a closely related genetic algorithm for a similar small-dimensional nonconvex optimization, with strong performance that further supports the effectiveness of such evolutionary approaches.}

To adopt DE, we start with a set of $P$ candidate random threshold vectors $\Pcal^{(0)} = \{\bt_p^{(0)}\}_{p=1}^P$ that satisfy the constraints, and iteratively update each member of the set (from $\Pcal^{(g-1)}=\{\bt_p^{(g-1)}\}$ to $\Pcal^{(g)} =\{\bt_p^{(g)}\}$) through the following mutation-crossover-selection procedure: 
\begin{itemize}
    \item (\textit{Mutation}) Find the current best solution $\bt^{(g-1)}_{\text{best}}$ out of $\Pcal^{(g-1)}$, and mutate it to form a ``donor'' $\bv = \bt^{(g-1)}_{\text{best}} + F_m(\bt_{r_1}^{(g-1)} - \bt_{r_2}^{(g-2)})$, where $r_1, r_2 \ne p$ are distinct indices and $F_m \sim \mathcal{U}(0.5, 1)$;
    \item (\textit{Crossover}) Generate a ``trial'' vector $\bu_p$ by inheriting each coordinate from either \(\bt_p^{(g-1)}\) or \(\bv\) with probability \(C_r\in[0,1]\), forcing at least one coordinate to come from \(\bv\);
    \item (\textit{Selection}) Update \(\bt^{(g)}_p \gets \bu_p\) if \(L(\bu_p;\bh)\le L(\bt^{(g-1)}_p;\bh)\); otherwise retain \(\bt^{(g)}_p \gets\bt^{(g-1)}_p\).
\end{itemize}
We enforce the monotonicity constraint during selection using feasibility-based rules \citep{lampinenConstraintHandlingApproach2002}, and select the final solution as the best one out of the final set.  
To implement this algorithm, we modify \texttt{differential\_evolution} function in the Python library \texttt{Scipy} \citep{virtanenSciPy10Fundamental2020} to our setting, while keeping the default hyperparameters of population size $P=15K$, mutation factor $F_m\sim \mathcal{U}(0.5, 1)$, and crossover probability $C_r=0.7$. Full algorithmic details and pseudocode are given in Section C of the supplementary material. 

In our simulations, we demonstrate that DE consistently achieves lower loss values than stepwise procedures, even when the solution thresholds are discretized, confirming its effectiveness for our problem.

\section{Simulations}\label{sec: simulations}

We evaluate the performance of the proposed method on simulated data with known optimal thresholds. We consider both joint continuous optimization \eqref{eq: opt-continuous}, based on differential evolution (DE), and greedy discrete optimization \eqref{eq: opt-discrete}, based on stepwise aggregation (SA) and stepwise splitting (SS). For each algorithm, we consider both $L_1$ and $L_2$ losses. For comparison, we use oracle, the known optimal thresholds, and principal amalgamation analysis (PAA) \citep{liPrincipalAmalgamationAnalysis2022}, a method from the compositional data literature. PAA optimizes the same objective as $L_2$ loss in \eqref{eq:loss_l2}, but replaces the 2-Wasserstein distance with Bray-Curtis dissimilarity and solves the resulting optimization problem using stepwise aggregation. Since PAA merges compositional variables regardless of their positions, we further impose the constraint to merge only neighboring variables to align with the histogram structures. When measuring the performance of each method, we report both estimated thresholds and the resulting value of the empirical loss. For PAA, we only report thresholds as loss values are not comparable due to the utilization of different distance measures.

To generate the data separated by $K^*$ predefined thresholds, we consider a mixture of uniform distributions for each sample $i=1, \dots, n$ with $n=200$:
\begin{equation*}
    M_i := w_{i0}U_{i0} + w_{i1}U_{i1} + \cdots + w_{iK^*}U_{iK^*},
\end{equation*}
where $w_{ij}$ are positive weights with $\sum_{j}w_{ij} = 1$, and $U_{ij}$ denotes the uniform distribution on the interval $I_{ij} = [c_{ij}, c_{i,j+1}]$, with $c_{i0} = 40$ and $c_{i, K^* + 1} = 400$ the lower and upper bounds, respectively. Each $M_i$ thus represents a histogram with thresholds $c_{ij}$, modeled by $c_{ij} = c_j + \varepsilon_{ij}$, $j=1,\ldots, K^*$, where $40 < c_1 < \cdots < c_{K^*} < 400$ are fixed base thresholds and $\varepsilon_{ij}$ are independent noise terms producing variability on the thresholds. In the noiseless case, the thresholds $\{c_j\}_{j=1}^{K^*}$ minimize both loss functions $L_1$ and $L_2$ evaluated with histograms $M_i$. Each empirical distribution $\hat M_i$ is constructed from 1000 observations drawn from the mixture distribution $M_i$. For the respective loss calculations, the oracle and DE utilize empirical quantiles of $\hat M_i$, while discrete methods (SA, SS, and PAA) first convert $\hat M_i$ into histograms with 180 bins, defined by cutoffs $40, 42, \ldots, 400$. For oracle, DE, SA, and SS, the grid size of $M=200$ is used for $L^2$ distance approximations in \eqref{eq: loss1} and \eqref{eq: loss2}. 

Throughout, we set $K^*=3$ with base thresholds $(c_1, c_2, c_3) = (70, 180, 250)$, {which induce four interval widths of 30, 110, 70, and 150 on $[40,400]$, reflecting heterogeneous window sizes.} We draw threshold perturbation $\varepsilon_{ij}$ i.i.d. from $N(0, \nu^2)$ with $\nu = 0, 5$, and 10, truncated to lie within $[-30, 30]$ to ensure the ordering of thresholds is preserved. For the weights $\bw_i = (w_{i0},\ldots, w_{iK^*})$, we consider two settings to explore the differences between the $L_1$ and $L_2$ loss functions:
\begin{itemize}
    \item \textbf{Setting 1}: $\bw_i\sim$ Dirichlet($20\cdot(0.3, 0.4, 0.2, 0.1)$)
    \item \textbf{Setting 2}: $(w_{i0}, w_{i1})\sim$ $0.7\cdot\text{Dirichlet}(5\cdot(0.5, 0.5))$ and $(w_{i2}, w_{i3}) = (0.2, 0.1) \mbox{[fixed]}$
\end{itemize}
Setting 1 represents a case where all four weights vary moderately between the samples, with optimal thresholds for both $L_1$ and $L_2$ losses expected to be $(70, 180, 250)$. Setting~2 keeps the last two weights constant across samples while only varying the first two. Optimization with $L_1$~\eqref{eq: loss1}, which preserves the overall shape of the distributions, is expected to recover $(70, 180, 250)$. For $L_2$~\eqref{eq: loss2}, which preserves pairwise distances, setting $K=2$ should yield $(70, 180)$, while $K=3$ should include a variable 3rd threshold across replications since the region $(180, 400)$ is non-variable across samples, making 250 non-informative for $L_2$.  Visualizations of the empirical quantile functions of simulated data in each setting are {presented in Figure~A1} of the supplementary material. {Under each setting, we generate $n=200$ distributions. In Section~A.3 of the supplementary material, we also consider $n=50, 100$ to reflect smaller sample sizes often encountered in clinical studies; the results are overall similar to $n=200$ and lead to the same conclusions.}

\begin{table}[t!]
    \centering
    \caption{Simulation results for Setting 1 with $n=200$. Thresholds and achieved loss values are averaged over 100 repetitions, with standard errors in parentheses. Methods compared include proposed joint optimization with differential evolution (DE), greedy methods based on stepwise aggregation (SA) and stepwise splitting (SS), and principal amalgamation analysis (PAA). Oracle refers to the base thresholds $(70, 180, 250)$. Bold highlights the lowest loss values for each loss function.}
    \label{tab: setting1}
    \resizebox{\columnwidth}{!}{%
    \begin{tabular}{ccccccccccc}
    \toprule
      & & \multicolumn{4}{c}{$L_1$} & \multicolumn{4}{c}{$L_2$} & \multirow{2.4}{*}{PAA} \\
    \cmidrule(lr){3-6} \cmidrule(lr){7-10}
      Noise &  & Oracle & DE & SA & SS & Oracle & DE & SA & SS &  \\
    \midrule
        \multirow{4}{*}{$\nu=0$}
           & $t_1$ & 70 & 70.2 (0.0) & {{70.0} (0.0)} & 79.9 (0.3) & 
                     70 & 70.3 (0.0) & {{70.0} (0.0)} & {70.0} (0.0) & {70.0 (0.0)} \\
          & $t_2$ & 180 & {180.1 (0.1)} & 180.1 (0.0) & 226.4 (1.2) & 
                    180 & 179.9 (0.1) & 180.0 (0.1) & 185.9 (2.2) & 180.0 (0.0)  \\ 
          & $t_3$ & 250 & 251.0 (0.1) & {250.1 (0.0)} & 251.8 (0.3) & 
                    250 & 252.2 (0.1) & 250.1 (0.0) & 261.4 (0.4) & 250.0 (0.0)   \\ 
          & Loss  & 6.42 (.03) & \textbf{6.35 (.03)} & 6.43 (.03) & 71.5 (1.2) & 
                    1.94 (.02) & \textbf{1.86 (.02)} & 1.95 (.02) & 10.0 (0.3) & --                 \\
        \midrule
         \multirow{4}{*}{$\nu=5$} 
            & $t_1$ & 70 & {73.5 (0.1)} & 76.1 (0.3) & 81.5 (0.3) & 
                      70 &  74.3 (0.1) & 76.0 (0.5) & 72.9 (0.2) & 74.9 (0.6) \\
           & $t_2$ & 180 & {179.9 (0.1)} & 180.3 (0.4) & 227.8 (1.3) & 
                     180 & 179.1 (0.1) & 179.8 (0.5) & 181.1 (2.5) & 178.5 (0.3)  \\
           & $t_3$ & 250 & {255.4 (0.1)} & 256.3 (0.5) & 257.5 (0.5) & 
                     250 & 256.0 (0.2) & 257.5 (0.6) & 260.6 (0.5) & 256.8 (0.5) \\   
           & Loss  & 17.1 (0.1) & \textbf{11.5 (0.1)} & 15.7 (0.4) & 82.3 (1.0) & 
                     5.88 (.07) & \textbf{3.59 (.04)} & 5.33 (.13) & 13.5 (0.3) &  --  \\
        \midrule
        \multirow{4}{*}{$\nu=10$}
            & $t_1$ & 70 & {74.1 (0.1)} & 78.5 (0.7) & 83.0 (0.2) & 
                      70 & 74.7 (0.2) & 77.4 (0.8) & 72.9 (0.2) & 67.6 (0.9) \\
           & $t_2$ & 180 & {178.3 (0.2)} & 176.6 (0.9) & 229.0 (0.6) & 
                     180 & 178.0 (0.2) & 176.1 (1.0) & 181.3 (2.7) & 170.7 (2.3) \\
           & $t_3$ & 250 & {257.9 (0.1)} & 263.0 (0.8) & 264.1 (0.3) & 
                     250 & 258.4 (0.2) & 262.4 (0.9) & 260.9 (0.6) & 251.2 (2.1)  \\  
           & Loss  & 32.9 (0.3) & \textbf{23.6 (0.1)} & 37.6 (0.9) & 97.1 (0.9) & 
                     11.3 (0.1) &  \textbf{8.03 (.07)} & 13.3 (0.3) & 20.4 (0.4) & --  \\
    \bottomrule
    \end{tabular}
    }
\end{table}

\cref{tab: setting1} presents the simulation results for Setting 1 with $K=3$ over 100 replications. DE consistently achieves the lowest empirical loss, outperforming the empirical losses at the oracle thresholds $(70, 180, 250)$ with similar or lower variance. SA attains empirical losses comparable to the oracle but shows higher variance under noisy conditions $\nu = 5$ and 10. SS underperforms in every scenario: its initial threshold selections (i) cluster around 230 under $L_1$ loss, introducing substantial bias relative to the oracle thresholds, and (ii) vary substantially around 180 under $L_2$ loss, resulting in unstable and suboptimal solutions. As expected, the methods achieving smaller losses, DE and SA, closely recover oracle thresholds under both $L_1$ and $L_2$ losses. The compositional method PAA closely captures the oracle thresholds under low noise ($\nu=0$ and 5) but becomes highly variable at $\nu=10$, exceeding the variance observed in SA.

\begin{table}[t!]
    \centering
    \caption{Simulation results for Setting 2 with $n=200$ empirical distributions and $K=3$ for $L_1$ loss and $K=2$ for $L_2$ loss and PAA. Thresholds and achieved loss values are averaged over 100 repetitions, with standard errors in parentheses. Oracle thresholds are $(70, 180, 250)$ for $L_1$ and $(70, 180)$ for $L_2$. Bold highlights the lowest loss values for each loss function.}
    \label{tab: setting2}
    \resizebox{\columnwidth}{!}{%
    \begin{tabular}{cccccccccccc}
    \toprule
      & & \multicolumn{4}{c}{$L_1$, $K=3$} & \multicolumn{4}{c}{$L_2$, $K=2$} & \multirow{2.4}{*}{PAA, $K=2$}\\
    \cmidrule(lr){3-6} \cmidrule(lr){7-10}
      Noise &  & Oracle & DE & SA & SS & Oracle & DE & SA & SS & \\
    \midrule
        \multirow{4}{*}{$\nu=0$} & $t_1$ & 70 & 70.1 (0.0) & 70.0 (0.0) & 68.0 (0.0) & 70 & 75.6 (0.5) & 70.0 (0.0) & 130.4 (0.1) & 70.0 (0.0) \\
          & $t_2$ & 180 & 179.8 (0.1) & 180.1 (0.1) & 227.9 (0.3) & 180 & 193.5 (1.3) & 179.9 (0.8) & 340.1 (11.9) & 180.1 (0.1) \\
          & $t_3$ & 250 & 250.9 (0.1) & 250.0 (0.0) & 252.1 (0.1) & -- & -- & -- & -- & -- \\
          & Loss  & 6.54 (.03) & \textbf{6.46 (.03)} & 6.56 (.03) & 39.9 (0.4) & 3.02 (.03) & \textbf{2.77 (.02)} & 3.49 (.14) & 10.7 (0.1) & -- \\
        \midrule
        \multirow{4}{*}{$\nu=5$} & $t_1$ & 70 & 74.6 (0.1) & 77.7 (0.4) & 74.3 (0.1) & 70 & 78.2 (0.2) & 78.0 (0.4) & 129.6 (0.1) & 64.2 (0.3) \\
          & $t_2$ & 180 & 178.5 (0.1) & 178.3 (0.5) & 227.5 (0.3) & 180 & 199.8 (0.4) & 179.9 (0.9) & 241.9 (16.6) & 79.3 (1.1) \\
          & $t_3$ & 250 & 253.8 (0.1) & 254.4 (0.5) & 256.9 (0.1) & -- & -- & -- & -- & -- \\
          & Loss  & 20.8 (0.2) & \textbf{12.9 (0.1)} & 18.3 (0.4) & 48.5 (0.5) & 10.8 (0.2) & \textbf{4.71 (.04)} & 8.12 (.22) & 15.5 (0.1) & -- \\
        \midrule
        \multirow{4}{*}{$\nu=10$} & $t_1$ & 70 & 76.1 (0.1) & 79.9 (0.8) & 77.3 (0.1) & 70 & 79.7 (0.2) & 81.1 (0.8) & 127.0 (0.1) & 58.0 (0.4) \\
          & $t_2$ & 180 & 175.1 (0.2) & 173.2 (1.0) & 227.1 (0.2) & 180 & 205.1 (0.4) & 176.3 (1.8) & 131.0 (14.4) & 81.8 (0.6) \\
          & $t_3$ & 250 & 255.3 (0.1) & 258.0 (0.8) & 261.4 (0.2) & -- & -- & -- & -- & -- \\
          & Loss  & 38.5 (0.4) & \textbf{26.7 (0.2)} & 44.1 (1.2) & 66.3 (0.5) & 19.6 (0.2) & \textbf{10.5 (0.1)} & 20.4 (0.5) & 28.2 (0.2) & -- \\
    \bottomrule
    \end{tabular}
    }
\end{table}

\begin{figure}[t]
    
    \centering
    \includegraphics[width=0.85\linewidth]{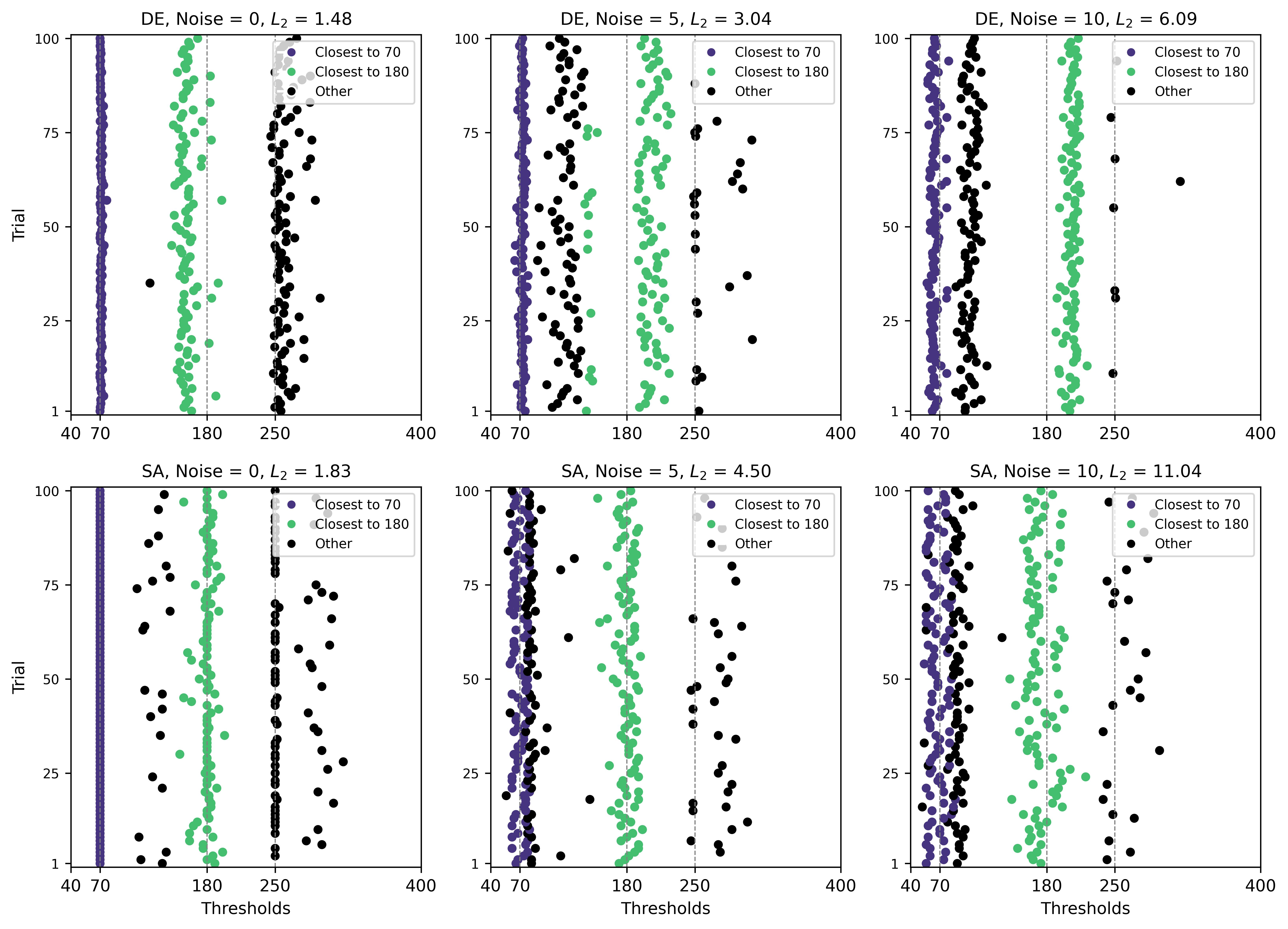}
    \caption{Results for simulation Setting 2 ($n=200$) with average $L_2$ loss for DE (top) and SA (bottom) under noise $\nu=0, 5, 10$. The $y$-axis represents 100 replications, with obtained thresholds from each replication plotted horizontally. Vertical dashed lines indicate the base thresholds $\bc=(70, 180, 250)$. Thresholds closest to 70 and 180 are colored in purple and green, respectively. The third threshold (black) is more variable and does not align with the base threshold $c_3 = 250$, which is non-informative for pairwise distances but informative for individual distributions.}
    \label{fig: setting2}
\end{figure}

For Setting 2, where the last two mixture weights are fixed, the results with $K=3$ under $L_1$ loss and $K=2$ under $L_2$ loss and PAA are reported in \cref{tab: setting2}. Again, DE achieves the lowest empirical losses, SA remains competitive, and SS continues to perform poorly across all scenarios. As expected, the well-performing methods, DE and SA, identify thresholds close to $(70, 180, 250)$ under the shape-preserving $L_1$ loss and close to $(70, 180)$ under the distance-preserving $L_2$ loss. Unexpectedly, PAA deteriorates significantly under noise $\nu=5$ and 10, though it identifies the oracle thresholds in the noiseless case. We attribute this behavior to PAA's reliance on the Bray-Curtis dissimilarity, which ignores the geometry of the underlying domain of distributions, unlike the 2-Wasserstein distance used in SA.
\cref{fig: setting2} further illustrates a scenario of selecting $K=3$ thresholds under $L_2$ loss for DE and SA. It shows that both DE and SA still identify thresholds near $(70, 180)$ while finding an extra threshold not necessarily related to $c_3=250$. In a few replications, DE selects thresholds not especially close to 180, yet they reduce $L_2$ loss by 50-70\% relative to the thresholds $(70, 180, 250)$, suggesting that this phenomenon stems from the random generation of empirical distributions $\hat M_i$.

In summary, DE consistently optimizes thresholds with high accuracy and stability. SA performs reasonably but exhibits higher variance under noisy cases. SS consistently underperforms despite its efficiency, and PAA, while effective in noiseless scenarios, deteriorates substantially under noisy conditions. Furthermore, DE is computationally more efficient than SA and PAA under our 180-bin setting, as stepwise methods require $O(J^2)$ searches: SA takes 5--10 times longer than DE under both $L_1$ and $L_2$ losses, and PAA has a similar runtime to SA under $L_1$. Consequently, DE stands out as the superior method, providing both accuracy and computational efficiency across all examined scenarios.

\section{Analysis of CGM data}\label{sec: real-data}

CGM data are commonly summarized using the consensus thresholds of 54, 70, 181, and 251 mg/dL \citep{battelinoContinuousGlucoseMonitoring2023}. 
Time spent below 54 mg/dL or between 54 and 69 mg/dL reflects severe or mild hypoglycemia, respectively, while time between 181 and 250 mg/dL or above 250 mg/dL indicates hyperglycemia and severe hyperglycemia, respectively. Prolonged exposure to hypo- or hyperglycemia significantly increases the risk of various health complications \citep{beckValidationTimeRange2018}. Thus, increasing the Time-in-Range (TIR) 70–180 mg/dL has become a standard clinical goal for diabetes management \citep{battelinoClinicalTargetsContinuous2019}. However, these thresholds are set primarily considering type 1 diabetes \citep{shahContinuousGlucoseMonitoring2019} and, to our knowledge, have not been assessed for optimality in a data-dependent manner. It thus remains unclear whether these thresholds are optimal or universally best for capturing clinically relevant signals in other populations, such as individuals without diabetes or with prediabetes.

In this section, we evaluate how the proposed data-driven thresholds compare with these consensus thresholds across three different types of analysis using CGM data from studies on diverse populations: (i) quality of representation of the underlying full CGM distribution in different populations; (ii) discriminatory power in distinguishing one population from the other; (iii) strength of associations with other clinical characteristics of interest. {We also consider naive thresholds as another baseline, defined by computing tertiles ($K=2$) or quintiles ($K=4$) of the pooled collection of all glucose readings across subjects, in the supplementary material (Section~B.4).} In all comparisons, we evaluate $K=4$ data-driven thresholds against the full consensus set $\{54, 70, 181, 251\}$ mg/dL, and $K=2$ data-driven thresholds against the standard TIR cutoffs $\{70, 181\}$ mg/dL. {Screeplots of the optimized loss as a function of $K$ for all real-data experiments are provided in Section~B.1 of the supplementary material, showing that $K=4$ thresholds are sufficient to capture distributional information}. We only present results from our differential evolution (DE) approach \eqref{eq: opt-continuous} given its superior performance demonstrated in simulations, using a grid size of $M=200$ for loss computations \eqref{eq: loss1} and \eqref{eq: loss2}. Since CGM readings are integer-valued, non-integer thresholds estimated by DE are rounded up, which does not affect the resulting summaries.

\subsection{Representational quality across populations}\label{sec:real-experim-separate}

In this section, we compare the representation quality of the thresholds in two distinct populations. Since the consensus thresholds were established primarily with type 1 diabetes in mind, we hypothesize that they should provide good representation for this population but may not be as effective for individuals without diabetes. To evaluate this, we consider two datasets.
The first dataset is from a multi-center prospective study by \citet{shahContinuousGlucoseMonitoring2019}, which aims to establish reference glucose ranges for individuals without diabetes using modern CGM devices. The second is from a randomized, multi-center clinical trial by \citet{brownSixMonthRandomizedMulticenter2019}, evaluating the efficacy and safety of an automated insulin pump system for individuals with type 1 diabetes. 
Both studies employ Dexcom G6 devices recording interstitial glucose every five minutes, offering high-resolution glucose time series for each participant. We included CGM profiles if they had $\geq $ 90\% of readings for $<$1 day, $ \geq 70\%$ for up to $2$ weeks, or $\geq 70\%$ of 14 days (9.8 days) for longer periods. 
On average, the group without diabetes is monitored for $1.3$ weeks, and the group with type 1 diabetes is monitored for $28.8$ weeks. Both datasets are publicly available from the JAEB Center for Health Research at \texttt{\url{https://public.jaeb.org}} with processing available through the Awesome-CGM repository \citep{xuIrinaStatsLabAwesomeCGMUpdated2024}.

We determine data-driven thresholds separately for the non-diabetes and type 1 diabetes datasets using the $L_1$ loss \eqref{eq: loss1}, which preserves distributional shapes and is well-suited for cohort-specific summaries. 
{In the supplementary material, we provide results with $L_2$ loss (Section B.2), which are qualitatively similar, and comparison with naive pooled-quantile thresholds, which underperform DE and are also inferior to consensus thresholds in type 1 diabetes (Section B.4).}

\begin{figure}
    \centering
    \includegraphics[width=\linewidth]{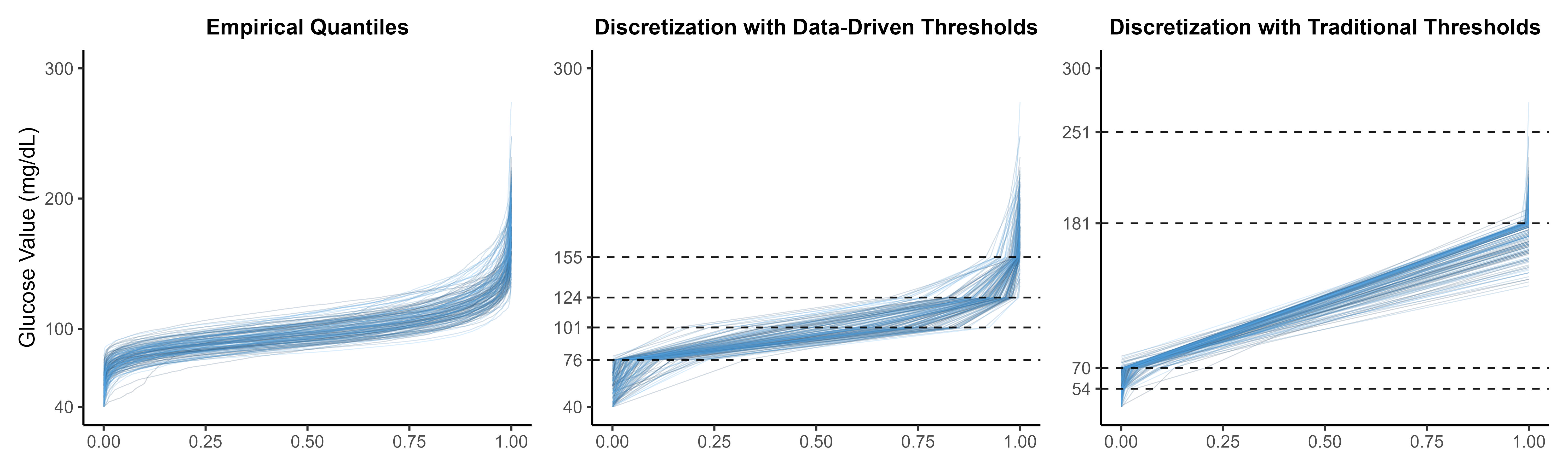}
    \caption{Quantile plots of glucose values for 168 individuals without diabetes from \citet{shahContinuousGlucoseMonitoring2019}, where each line represents a single subject. Empirical quantiles are shown in the left panel, while the middle and right panels display piecewise linearized quantiles using data-driven thresholds ($L_1$ loss, $K = 4$) and the traditional thresholds, respectively, with thresholds indicated by horizontal lines.}
    \label{fig: shah_quantiles}
\end{figure}

For the data on $n=168$ individuals without diabetes \citep{shahContinuousGlucoseMonitoring2019}, the four consensus thresholds $\{54, 70, 181, 251\}$ mg/dL yield $L_1 = 655.9$. In contrast, our DE approach with $K=4$ attains thresholds at $\{76, 101, 124, 155\}$ mg/dL, which vastly reduces $L_1$ to $16.6$, a 97\% reduction. \cref{fig: shah_quantiles} illustrates how these narrower cutoffs more accurately capture the glucose distributions of individuals without diabetes via piecewise linearized quantiles, following the perspective in \cref{sec: quantile-linearization}. Here, glucose readings rarely exceed 180 mg/dL, making the large consensus cutoffs of 181 and 251 mg/dL largely uninformative. Even with $K=2$, DE finds thresholds at $\{72, 128\}$ mg/dL achieving $L_1 = 88.9$, still an 86\% reduction relative to the four consensus thresholds. These results reflect the typically narrower glycemic distributions of individuals without diabetes, as reported by \citet{shahContinuousGlucoseMonitoring2019}.

For the data on $n=168$ individuals with type 1 diabetes \citep{brownSixMonthRandomizedMulticenter2019}, the four consensus thresholds yield $L_1 = 159.9$, a smaller average discrepancy than the data on individuals without diabetes. Still, DE with $K=4$ attains thresholds at $\{85, 172, 233, 302\}$ mg/dL, which reduces $L_1$ to 41.2, a 74\% reduction. Notably, the first three data-driven cutoffs roughly align with the conventional 70, 181, and 251 mg/dL, while the newly identified 302 mg/dL suggests finer partitioning in the severe hyperglycemic region, which is frequently reached or exceeded in type 1 diabetes (see Figure S4 in the supplementary material for their empirical quantiles). With $K=2$, DE similarly finds higher thresholds at $\{211, 289\}$ mg/dL achieving $L_1={398.2}$, a {68}\% reduction from the $L_1 = 1236.0$ at the standard TIR pair $\{70, 181\}$ mg/dL. These findings illustrate that higher cutoffs better capture glucose distributions for individuals with type 1 diabetes.

Despite this appeal for higher thresholds, monitoring hypoglycemia (below 70 mg/dL) or increasing the TIR of 70--180 mg/dL remains crucial for type 1 diabetes management. To address this, we also apply our semi-supervised approach \eqref{eq: opt-semisuper} to the same data, fixing $\bt_{\text{fix}}=\{70, 181\}$ mg/dL and optimizing two additional thresholds. Under this scheme, DE identifies thresholds $\{70, 181, 241, 306\}$ mg/dL, where additional thresholds split higher glycemic regions similarly. Although they yield a slightly higher loss $L_1 = 63.8$ than the fully data-driven solution, this still represents a 60\% improvement from the four consensus thresholds. Thus, it offers more informative summaries for type 1 diabetes while preserving the standard TIR range. 

Overall, these findings demonstrate that the fixed consensus thresholds $\{54, 70, 181, 251\}$ mg/dL do not optimally capture the distinct glycemic patterns of individuals without diabetes versus those with type 1 diabetes. By contrast, our data-driven thresholds provide {cohort-adaptive thresholds by optimizing the distribution-preservation $L_1$ loss}, suggesting narrower cutoffs for individuals without diabetes and higher thresholds for those with type 1 diabetes. The semi-supervised approach offers a middle ground for type 1 diabetes, preserving key clinical benchmarks while also identifying higher additional thresholds, thereby enhancing the information captured by TIR summaries.

\subsection{Discriminative power in group classification}

To compare the discriminatory power of the data-driven and consensus thresholds, we combine the two datasets from Section~\ref{sec:real-experim-separate} ($n=336$ in total). We generate the data-driven thresholds using the combined dataset with the $L_2$ loss \eqref{eq: loss2} to preserve between-subject distances, an objective widely used in classification and clustering tasks for discriminating between subpopulations (the $L_1$ loss does not necessarily target discrimination; see Section B.2 of the supplementary material). We then evaluate both the resulting data-driven and consensus thresholds for their ability to effectively discriminate between individuals with type 1 diabetes and those without diabetes.

\begin{figure}[t]
    \centering
    \includegraphics[width=0.8\linewidth]{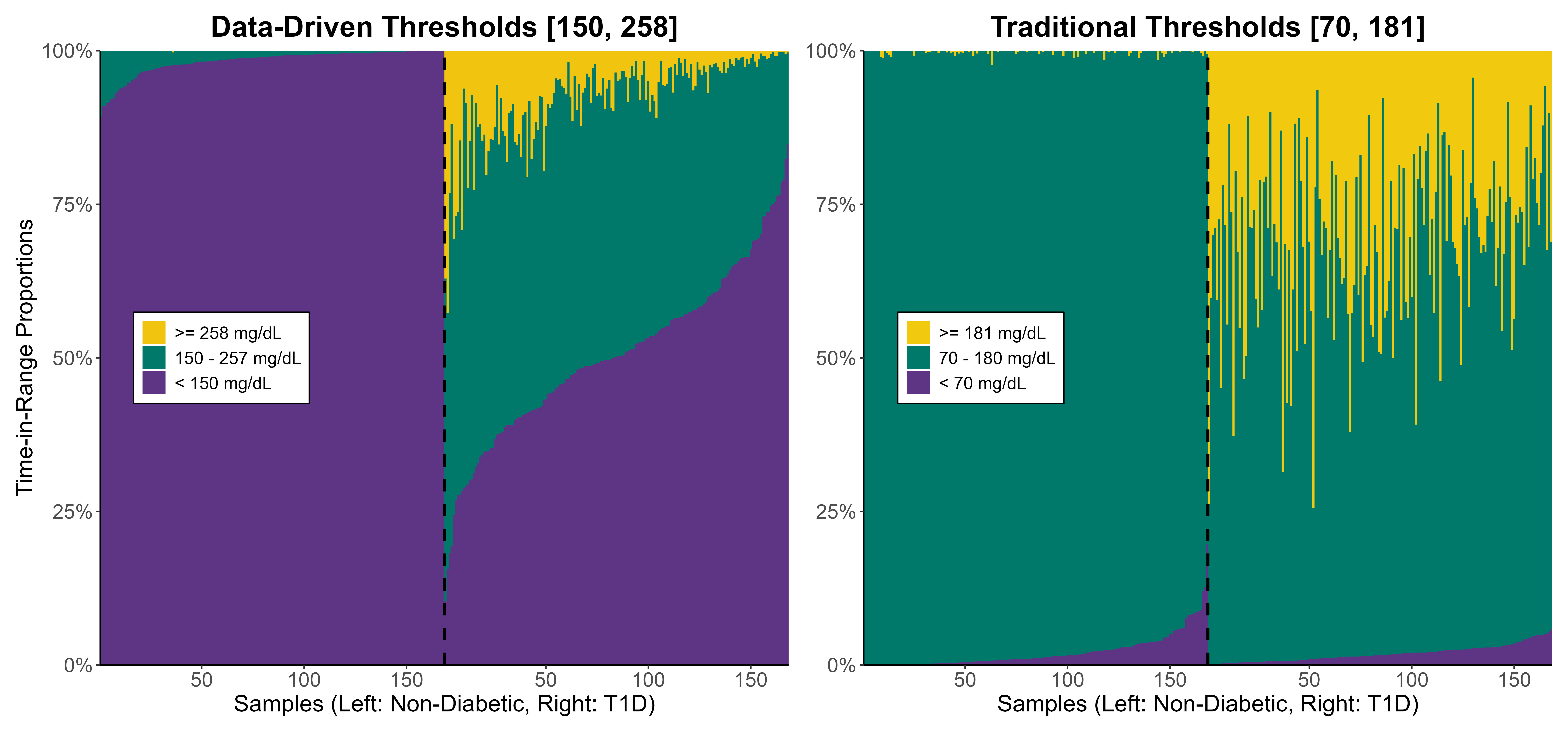}
    \caption{Compositional bar plots showing TIR proportions derived by data-driven thresholds $\{150, 258\}$ mg/dL ($L_2$ loss; left) and traditional thresholds $\{70, 181\}$ mg/dL (right). Individuals without diabetes and individuals with type 1 diabetes are separated by vertical black lines, with samples arranged by time below range within each group.}
    \label{fig: TIR barplots}
\end{figure}

For conciseness, we only illustrate results with $K=2$ thresholds, as $K=4$ thresholds yield similar interpretations; see Section B.2 (supplementary material). With $K=2$ thresholds, the consensus pair $\{70, 181\}$ mg/dL yields $L_2 = 111.6$. In contrast, DE identifies thresholds at $\{150, 258\}$ mg/dL, which reduces $L_2$ to 22.2, an 80\% improvement. \cref{fig: TIR barplots} compares the resulting TIR summaries via compositional bar plots for both pairs of thresholds. As depicted, every interval defined by $\{150, 258\}$ mg/dL clearly separates the two groups, whereas TIR $<70$ mg/dL fails, and TIR 70--180 mg/dL provides weaker separation than the data-driven thresholds. We also fit downstream logistic regression using each TIR proportion, with results shown in \cref{tab: combined_data}. Under data-driven thresholds, individuals with type 1 diabetes are identified by spending less than 86.6\% TIR $< 150$ mg/dL, more than 12.8\% TIR 150--257 mg/dL, and more than 0.2\% TIR $\ge 258$ mg/dL. 
In contrast, the consensus thresholds fail to discriminate the two groups via TIR $<70$ mg/dL, while achieving weaker discrimination than data-driven ranges with TIR 70--180 mg/dL and perfect discrimination with TIR $\ge 181$ mg/dL.
{Naive thresholds achieve overall better discrimination than consensus thresholds but are inferior to DE; see Section~B.4 of the supplementary material.}
Consequently, our $L_2$-optimized thresholds demonstrate improved discrimination over consensus thresholds, highlighting their potential for classification or clustering with interpretable TIR summaries in mixed-population studies.

\begin{table}[t]

\centering
\caption{Univariate logistic regression results for each TIR proportion under the data-driven (DE) thresholds $\{150, 258\}$ mg/dL versus consensus thresholds $\{70, 181\}$ mg/dL. ``Accuracy (\%)'' reports the classification accuracy, and ``Decision Boundary'' the estimated TIR cutoff that discriminates between individuals without diabetes and those with type 1 diabetes.}
\resizebox{\columnwidth}{!}{%
\begin{tabular}{lcclcc}
\toprule

\multicolumn{3}{c}{DE} & \multicolumn{3}{c}{Consensus} \\
\cmidrule(lr){1-3}\cmidrule(lr){4-6}
Ranges (mg/dL) & Accuracy (\%) & Decision boundary & Ranges (mg/dL) & Accuracy (\%) & Decision boundary \\
\midrule
TIR $<150$      & 100.0 & 0.866 & TIR $<70$      &  47.0 & 0.020 \\
TIR 150--257    & 100.0 & 0.128 & TIR 70--180    &  97.9 & 0.893 \\
TIR $\ge 258$   &  99.1 & 0.002 & TIR $\ge 181$  & 100.0 & 0.033 \\
\bottomrule
\end{tabular}
}
\label{tab: combined_data}
\end{table}

\subsection{Strength of association with other clinical outcomes}\label{sec:ai-readi}

In this section, we evaluate the strength of association between the threshold-based glycemic summaries and lipid profiles, specifically triglycerides (TG), HDL cholesterol (HDL-C), LDL cholesterol (LDL-C), total cholesterol (Total-C), which together constitute a major risk factor for cardiovascular disease and are closely linked with diabetes pathophysiology \citep{zhaoTriglycerideIndependentPredictor2019}. We also evaluate association with TG/HDL-C ratio, a surrogate marker of insulin resistance \citep{chenTriglycerideHighdensityLipoprotein2022}. For this goal, we consider the AI-READI dataset \citep{ai-readiconsortiumAIREADIRethinkingAI2024}, a multi-site study with on-site laboratory and survey assessments followed by $\sim$10 days of masked Dexcom G6 CGM recording. Participants include individuals without diabetes, those with prediabetes, and those with type 2 diabetes, determined by their medical records. To minimize confounding from treatment effects, our analysis focuses on $n = 573$ individuals not treated with insulin or oral medications (HbA1c 4.5–7.4\%), encompassing people without diabetes (313 individuals have HbA1c $< 5.7$\%) and those with prediabetes or lifestyle-controlled type 2 diabetes.

To determine data-driven thresholds, we use the $L_2$ loss to preserve between-subject glycemic variability. With $K=2$ cutoffs, DE identifies thresholds $\{96, 170\}$ mg/dL, which reduce $L_2$ by 56\% relative to the consensus pair $\{70, 181\}$ mg/dL. The upper threshold at 170 mg/dL captures hyperglycemia similarly to 181 mg/dL but more tightly. The lower threshold at 96 mg/dL aligns with impaired fasting glucose beginning at 100 mg/dL for prediabetes \citep{rooneyGlobalPrevalencePrediabetes2023}, positioning TIR $<96$ mg/dL as another discriminatory region for better glycemic control. With $K=4$, DE obtains $\{90, 128, 172, 232\}$ mg/dL that improves $L_2$ by 95\% over the consensus thresholds $\{54, 70, 181, 251\}$ mg/dL; thresholds at 90 and 172 mg/dL admit similar interpretation to the $K=2$ case, while 128 and 232 mg/dL provide a finer subdivision of near-normoglycemic and more severe hyperglycemic regions.

We compare how the new data-driven thresholds relate to each of the lipid profiles relative to consensus thresholds.
Table~\ref{tab:kendall_tau} reports the Kendall's $\tau$ between each of the resulting proportions and lipid profile. With $K=2$ thresholds, the consensus TIR 70--180 is largely uninformative, whereas the DE's 96--169 band shows modest but significant positive associations with TG and TG/HDL-C ($p < 0.05$). DE's lower band TIR $<96$ exhibits stronger, sign-consistent associations than TIR $<70$ (positive with HDL-C; negative with TG and TG/HDL-C). With $K=4$ thresholds, the middle data-driven cutoff at 128 mg/dL yields a clear sign change: below 128 mg/dL, the TIR proportions correlate positively to HDL-C and negatively to TG and TG/HDL-C, and the pattern reverses above 128 mg/dL. Consensus thresholds fail to capture this pattern, suggesting that the 70--180 mg/dL band is too coarse to capture lipid-relevant glycemic structure in this population. LDL-C and Total-C show no dependence on any TIR component, consistent with reports that CGM metrics primarily relate to TG and HDL-C rather than LDL-C and Total-C \citep{salsa-casteloAssociationGlycemicVariability2024}.

\begin{table}[!t]
\centering
\caption{Kendall's $\tau$ correlation between TIR proportions and lipid variables: HDL cholesterol (HDL-C), LDL-cholesterol (LDL-C), total cholesterol (Total-C), triglycerides (TG), and TG/HDL-C ratio. Significant $p$-values under the null hypothesis of $\tau = 0$ are indicated alongside.}
\label{tab:kendall_tau}
\footnotesize
\begin{tabular}{lllrrrrr}
\toprule
$K$ & Method & Range (mg/dL) & \multicolumn{1}{c}{HDL-C} & \multicolumn{1}{c}{LDL-C} & \multicolumn{1}{c}{Total-C} & \multicolumn{1}{c}{TG} & \multicolumn{1}{c}{TG/HDL-C} \\
\midrule
\multirow{6.6}{*}{$K=2$} & \multirow{3}{*}{Consensus} & TIR $<70$ & 0.060$^*$ & 0.002$\:\,$ & -0.000$\:\,$ & -0.086$^\dagger$ & -0.091$^\dagger$ \\
 &  & TIR 70--180 & 0.012$\:\,$ & -0.005$\:\,$ & -0.015$\:\,$ & -0.024$\:\,$ & -0.021$\:\,$ \\
 &  & TIR $\ge181$ & -0.033$\:\,$ & -0.002$\:\,$ & 0.018$\:\,$ & 0.079$^\dagger$ & 0.071$^*$ \\
\cmidrule(lr){2-8}
 & \multirow{3}{*}{DE} & TIR $<96$ & 0.088$^\dagger$ & -0.013$\:\,$ & -0.017$\:\,$ & -0.113$^\ddagger$ & -0.119$^\ddagger$ \\
 &  & TIR 96--169 & -0.027$\:\,$ & -0.002$\:\,$ & 0.005$\:\,$ & 0.058$^*$ & 0.057$^*$ \\
 &  & TIR $\ge170$ & -0.042$\:\,$ & -0.006$\:\,$ & 0.013$\:\,$ & 0.083$^\dagger$ & 0.079$^\dagger$ \\
\midrule
\multirow{10.6}{*}{$K=4$} & \multirow{5}{*}{Consensus} & TIR $<54$ & 0.061$\:\,$ & 0.030$\:\,$ & 0.041$\:\,$ & -0.026$\:\,$ & -0.042$\:\,$ \\
 &  & TIR 54--69 & 0.063$^*$ & -0.008$\:\,$ & -0.009$\:\,$ & -0.098$^\ddagger$ & -0.101$^\ddagger$ \\
 &  & TIR 70--180 & 0.012$\:\,$ & -0.005$\:\,$ & -0.015$\:\,$ & -0.024$\:\,$ & -0.021$\:\,$ \\
 &  & TIR 181--250 & -0.035$\:\,$ & -0.003$\:\,$ & 0.018$\:\,$ & 0.080$^\dagger$ & 0.073$^\dagger$ \\
 &  & TIR $\ge251$ & 0.036$\:\,$ & 0.021$\:\,$ & 0.038$\:\,$ & 0.022$\:\,$ & 0.001$\:\,$ \\
\cmidrule(lr){2-8}
 & \multirow{5}{*}{DE} & TIR $<90$ & 0.081$^\dagger$ & -0.011$\:\,$ & -0.015$\:\,$ & -0.107$^\ddagger$ & -0.112$^\ddagger$ \\
 &  & TIR 90--127 & 0.125$^\ddagger$ & 0.001$\:\,$ & 0.020$\:\,$ & -0.083$^\dagger$ & -0.113$^\ddagger$ \\
 &  & TIR 128--171 & -0.143$^\ddagger$ & 0.003$\:\,$ & -0.006$\:\,$ & 0.130$^\ddagger$ & 0.156$^\ddagger$ \\
 &  & TIR 172--231 & -0.045$\:\,$ & -0.004$\:\,$ & 0.013$\:\,$ & 0.084$^\dagger$ & 0.080$^\dagger$ \\
 &  & TIR $\ge232$ & 0.020$\:\,$ & 0.033$\:\,$ & 0.059$\:\,$ & 0.056$\:\,$ & 0.032$\:\,$ \\
\bottomrule
\end{tabular}
\begin{tablenotes}
\scriptsize
\item $\quad\ \ $ * $p<0.05$, † $p<0.01$, ‡ $p<0.001$
\end{tablenotes}
\end{table}

To further compare the strength of associations between consensus and data-driven thresholds while accounting for the entire composition, we fit linear models with each lipid profile as an outcome and compositional TIR predictors. Due to the inherent multicollinearity, we omit the highest TIR bin from the set of predictors; the resulting model is equivalent to the intercept-free, identifiable model for compositional predictors in \citet{liItsAllRelative2023}. {As a full-distribution benchmark, we also include Wasserstein regression (WR) \citep{chenWassersteinRegression2023}, which uses the entire distribution as a predictor.}
TG and TG/HDL-C are log-transformed to address skewness, and LDL-C and Total-C are excluded since the $F$-test shows no significant effects in all cases. Table~\ref{tab:lm_results} shows that DE consistently improves model fit, achieving higher $R^2$ and lower AIC (often by $>10$) compared with consensus thresholds. Clarke’s observation-wise likelihood-ratio test for non-nested {linear} models \citep{clarkeSimpleDistributionFreeTest2007} favors DE at significance level $\alpha=0.05$ throughout. Notably, for HDL-C, the performance of data-driven thresholds at $K=4$ is significantly better than both consensus thresholds and data-driven thresholds at $K=2$. This is aligned with the substantially stronger Kendall’s $\tau$ captured around the threshold at 128 mg/dL in Table~\ref{tab:kendall_tau}, with the reverse association pattern between TIR 90--127 and TIR 128--171 bins, compared to non-significant $\tau$ with wider ranges of TIR 70--180 and TIR 96--169. 
{WR attains the highest $R^2$ values as expected from using the full distributions, and DE with $K=4$ closely tracks WR, consistent with the screeplot in Section~B.1 of the supplementary material showing that $K=4$ thresholds capture most pairwise distributional distances.}
While the overall predictive power is modest in alignment with prior findings \citep{salsa-casteloAssociationGlycemicVariability2024}, the associations are significant with data-driven thresholds exhibiting consistently stronger associations and stronger predictive performance, suggesting they lead to higher statistical power to detect associations between CGM profiles and other clinical variables. 
{Section~B.4 of the supplementary material reports comparisons with naive thresholds, which perform competitively or sometimes better than DE for some lipid outcomes. However, naive thresholds are suboptimal in distributional representation and fail to improve on the consensus thresholds in the additional HbA1c prediction reported there, whereas DE remains consistently strong and close to WR.}

\begin{table}[t]
\centering
\caption{Comparison of linear model fits with TIR compositional predictors based on data-driven (DE) and consensus (CS) thresholds. {Wasserstein regression (WR), which uses the full distribution as a predictor, is included as a benchmark.} $\Delta$AIC denotes the difference AIC$_{\text{CS}}$ - AIC$_{\text{DE}}$. Significant $p$-values from Clarke's test for non-nested model comparison are indicated with $\Delta$AIC, confirming that DE significantly outperforms CS in all cases.}
\label{tab:lm_results}
\small
\begin{tabular}{lrrrrrrr}
\toprule
 & \multicolumn{3}{c}{$K=2$} & \multicolumn{3}{c}{$K=4$} & \multicolumn{1}{c}{WR} \\
\cmidrule(lr){2-4} \cmidrule(lr){5-7} \cmidrule(lr){8-8} Response & \multicolumn{1}{c}{$R^2_{\text{DE}}$} & \multicolumn{1}{c}{$R^2_{\text{CS}}$} & \multicolumn{1}{c}{$\Delta$AIC} & \multicolumn{1}{c}{$R^2_{\text{DE}}$} & \multicolumn{1}{c}{$R^2_{\text{CS}}$} & \multicolumn{1}{c}{$\Delta$AIC} & \multicolumn{1}{c}{{$R^2$}} \\
\midrule
HDL-C & 0.018 & 0.011 & 4.2$^\ddagger$ & 0.041 & 0.014 & 16.2$^\ddagger$ & {0.041} \\
TG & 0.049 & 0.025 & 14.5$^\ddagger$ & 0.052 & 0.040 & 7.5$^*$ & {0.054} \\
TG/HDL-C & 0.050 & 0.026 & 14.3$^\ddagger$ & 0.060 & 0.040 & 11.8$^\dagger$ & {0.067} \\
\bottomrule
\end{tabular}
\begin{tablenotes}
\footnotesize
\item \hspace{5em} * $p<0.05$, † $p<0.01$, ‡ $p<0.001$
\end{tablenotes}
\end{table}

\section{Discussions}

In this paper, we propose a new framework for data-driven thresholds to optimize wearable device data summaries, taking CGM data as our primary motivating example. By incorporating the Wasserstein distance into optimality measures inspired by amalgamation in compositional data analysis, we formulate threshold selection as optimal piecewise-linear approximations of quantiles. {This unsupervised formulation targets data-adaptive thresholds that preserve richer distributional information than pre-fixed threshold-based summaries while remaining compatible with the same downstream inferential pipelines.} We develop two algorithmic approaches, stepwise algorithms and a joint optimization with differential evolution (DE), {and simulations demonstrate the superior performance of DE. Analyses of CGM datasets reveal that $K=4$ data-driven thresholds often suffice to capture most distributional information, and that these optimal} thresholds differ across populations, improve discrimination performance, and exhibit stronger associations with clinical variables than the consensus thresholds, highlighting the potential for refining clinical applications.
The semi-supervised extension allows for fixing some thresholds of practical importance, balancing domain knowledge and distributional structures.
{The Python code and the R package \texttt{OptiThresholdR} are publicly available at: \texttt{\url{https://github.com/IrinaStatsLab/OptiThresholds}}}.

Our approach may inform the choice of appropriate thresholds in clinical research using wearable devices, particularly in populations where fixed thresholds may fail to capture meaningful patterns, such as in early-stage type 2 diabetes (\cref{sec:ai-readi}). For future applications, $L_1$-optimized thresholds can characterize population-level patterns in large cohorts, as in the CGM study of over 7000 individuals without diabetes \citep{keshetCGMapCharacterizingContinuous2023}. $L_2$-optimized thresholds could support clinical trials and outcome prediction, where {use of interpretable TIR endpoints with valid statistical inference is particularly crucial. More broadly, because our framework is unsupervised, it is also well suited to epidemiologic studies with multiple outcomes, where a single cohort-adaptive summary can be consistently used across multiple downstream analyses. Furthermore, while we focus on amalgamating histograms of integer-valued wearable device measurements, such as CGM data, our method naturally extends to general continuous distributional data via the continuous relaxation in \cref{sec: continuous-relaxation}.}

One limitation of our unsupervised thresholds is that, while optimized to represent underlying distributional signals from wearable measurements, they are not directly linked to disease or health-related risks. Studying how these data-driven thresholds---such as the hyperglycemic cutoffs identified in the type 1 diabetes dataset---relate to health outcomes will require prospective validation. Although prospective CGM studies remain limited, their expansion will provide crucial opportunities to test whether data-driven thresholds improve risk prediction beyond fixed consensus ranges.  {As linked outcome data becomes more available, an important direction for future research is the development of supervised threshold selection methods with valid post-selection inference guarantees. Such advancements would allow researchers to optimize thresholds for specific clinical endpoints while maintaining statistical validity on the full dataset, eliminating the need for power-reducing sample splitting.}

{Furthermore, as multimodal wearable data become increasingly prominent (e.g., concurrent accelerometry and heart rate measurements), extending our framework to multivariate wearable signals would be promising. In principle, placing thresholds along each coordinate could summarize multivariate distributions by the time spent in the resulting hyperrectangular regions. However, the resulting optimization is highly nontrivial due to the increasing number of optimization variables and the computational demands of the Wasserstein distance in higher dimensions \citep{peyreComputationalOptimalTransport2019}, making it an interesting direction for future work.}

\section*{Acknowledgment}
The source of the \citet{brownSixMonthRandomizedMulticenter2019} and \citet{shahContinuousGlucoseMonitoring2019} datasets is the JAEB Center for Health Research. The analyses, content, and conclusions presented herein are solely the responsibility of the authors and have not been reviewed or approved by the JAEB Center for Health Research. This research was supported by NIH R01HL172785.

\bibliographystyle{chicago}
\bibliography{JunyoungRef}

\newpage
\appendix
\bigskip

\setcounter{figure}{0}
\setcounter{table}{0}

\renewcommand{\thetable}{S\arabic{table}}  
\renewcommand{\thefigure}{S\arabic{figure}}
\renewcommand{\thealgorithm}{S\arabic{algorithm}}

\begin{center}
\LARGE{
Supplementary material for ``Beyond fixed thresholds: optimizing summaries of wearable device data via piecewise linearization of quantile functions'' by
}
\vskip 1em
\large{Junyoung Park, Neo Kok, and Irina Gaynanova}\\[1em]
\large{Department of Biostatistics, University of Michigan, Ann Arbor, Michigan, USA}
\end{center}
\bigskip

\begin{abstract}
    \noindent This supplementary material provides additional details about the experiments and implementations. Section~\ref{sec: supplement-simul} offers supplemental information on simulation settings and additional results. Section~\ref{sec: supplement-real} presents additional results from real-data experiments using CGM data. In Section~\ref{sec:supp-DE}, we provide the detailed implementation of the differential evolution algorithm for our optimal threshold estimation.
\end{abstract}

\section{Supplementary information for simulations}\label{sec: supplement-simul}

\subsection{Visualizations of the simulated empirical distributions}

\begin{figure}[htbp]
    \centering
    \includegraphics[width=0.9\linewidth]{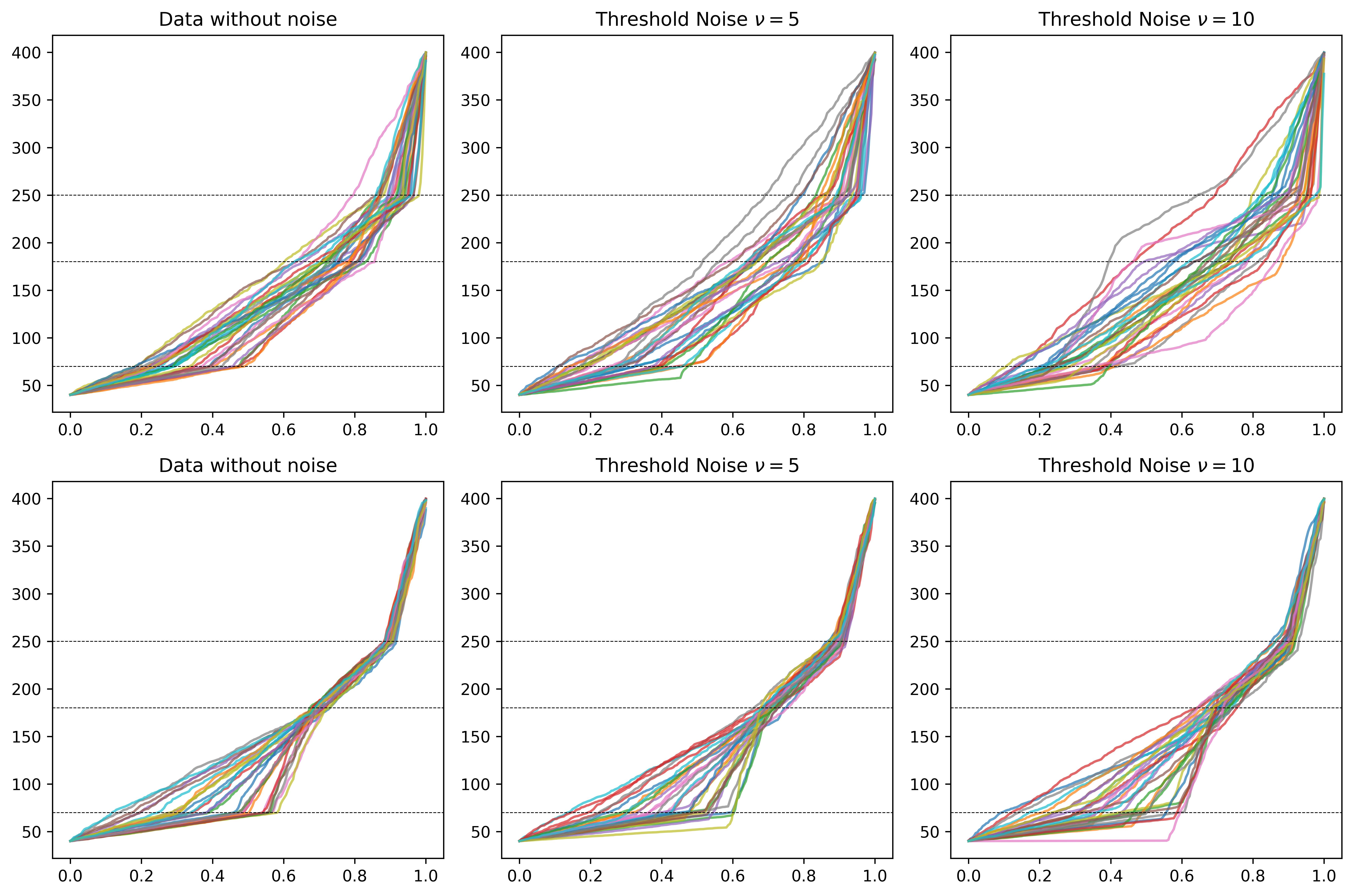}
    \caption{Visualization of the 30 empirical distributions generated for Setting 1 (top) and Setting 2 (bottom) with noise levels $\nu=0, 5,$ and 10. Base thresholds $\bc = (70, 180, 250)$ are shown as the dashed lines.}
    \label{fig: data-visual-sets1and2}
\end{figure}

We visualize the generated empirical quantiles $\hat M_i$ for simulation Settings 1 and 2 to illustrate the structures of simulated data. To simplify graphics, we depict only $n=30$ empirical distributions under each setting with noise levels $\nu=0, 5$, and 10 in Figure \ref{fig: data-visual-sets1and2}. The distributions show piecewise linear patterns with vertices around the base thresholds $\bc = (70, 180, 250)$, depicted as horizontal dashed lines. In the bottom figures on Setting 2, the variability of quantiles in the region $[180, 400]$ is tiny, representing that the pairwise distances are mostly computed along the region $[40, 180]$.

\subsection{Auxiliary simulation results for Setting 2}

\begin{figure}[htbp]
    \centering
    \includegraphics[width=0.9\linewidth]{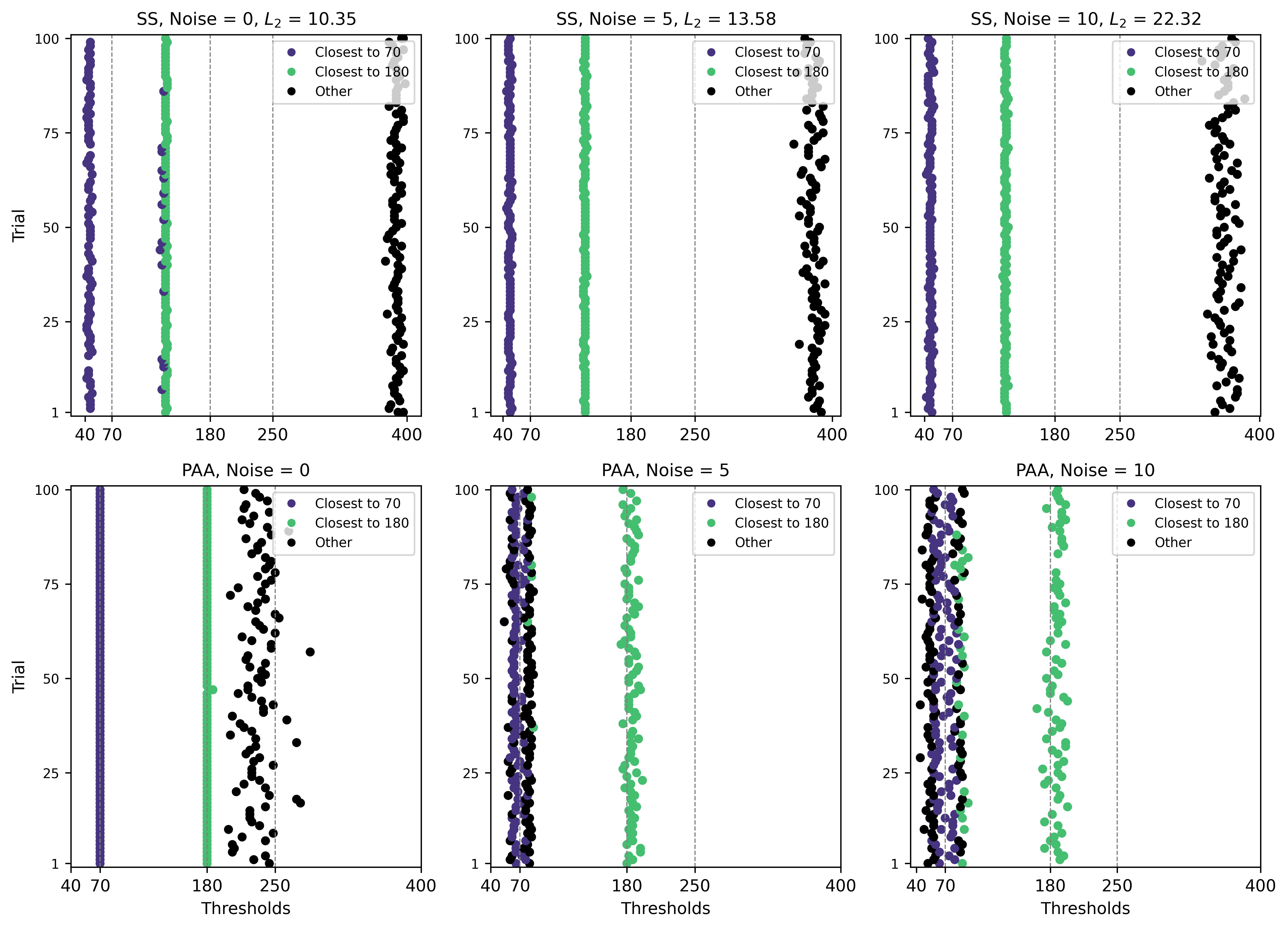}
    \caption{Simulation results under Setting 2 with $K=3$ for PAA and SS ($L_2$ loss). The $y$-axis represents 100 replications, and the thresholds obtained in each replication are plotted horizontally. Vertical dashed lines illustrate the base thresholds $\bc=(70, 180, 250)$. Thresholds closest to 70 and 180 are colored in purple and green, respectively. $L_2$ loss values are averaged and divided by $10^3$.}
    \label{fig: supp-simul-set2-K3}
\end{figure}

\begin{figure}[htbp]
    \centering
    \includegraphics[width=0.9\linewidth]{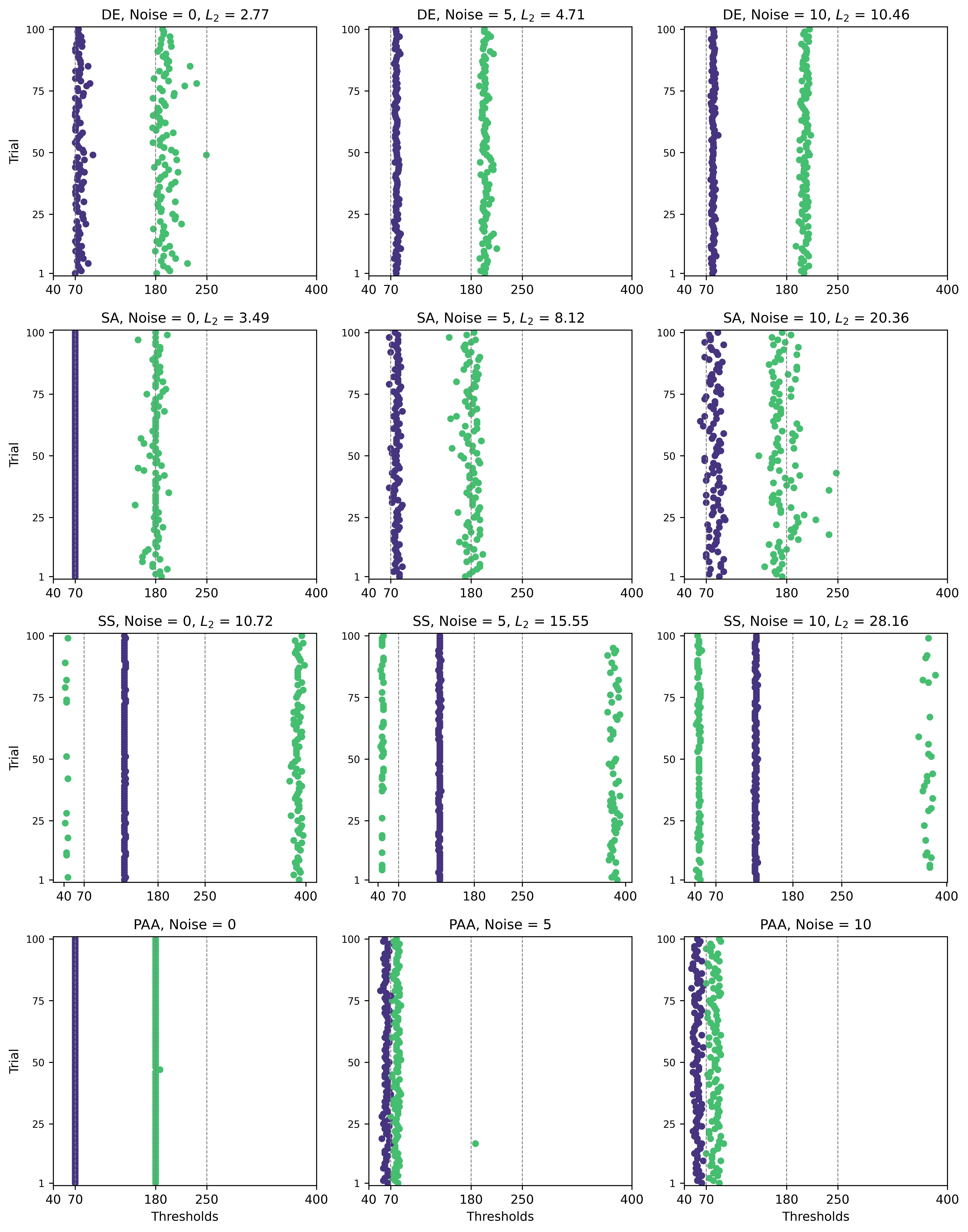}
    \caption{Simulation results under Setting 2 with $K=2$ and $L_2$ loss for  DE, SA, and SS, while PAA is also presented at the bottom. The $y$-axis represents 100 replications, and the thresholds obtained in each replication are plotted horizontally. Vertical dashed lines illustrate the base thresholds $\bc=(70, 180)$. Thresholds closest to 70 and 180 are colored in purple and green, respectively. $L_2$ loss values are averaged and divided by $10^3$.
    }
    \label{fig: supp-simul-set2-K2}
\end{figure}

Figures \ref{fig: supp-simul-set2-K3} and \ref{fig: supp-simul-set2-K2} present additional simulation results under Setting 2 with thresholds obtained using the $L_2$ loss for $K=3$ and $K=2$, respectively. \cref{fig: supp-simul-set2-K3} depicts the underperforming methods, SS and PAA, while \cref{fig: supp-simul-set2-K2} includes all methods considered in the paper: DE, SA, SS, and PAA. PAA with $K=3$ shows an interesting trend, where it often detects thresholds close to $180$ even in the noisy cases of $\nu=5$ and 10, though it also often collapses to the thresholds near 70 as in the case $K=2$.

\subsection{{Additional simulation results with smaller sample size $n$}}

{To assess robustness to smaller sample sizes commonly encountered in wearable-device studies, we repeat the simulations under the same data-generating settings (Settings 1 and 2) with $n=50$ and $n=100$, in addition to the main results for $n=200$. Results are summarized in Tables \ref{tab: setting1_n50}--\ref{tab: setting2_n100}.
The qualitative conclusions remain unchanged across sample sizes. 
In Setting 1, DE consistently attains the smallest empirical losses (even smaller than the oracle), SA remains competitive with higher variability in determined thresholds, SS remains suboptimal, and PAA becomes less stable as noise increases. In Setting 2, the DE and SA still recover thresholds near $(70,180,250)$ under $L_1$ and near $(70,180)$ under $L_2$, while SS remains suboptimal, and PAA deteriorates significantly when selecting $K=2$ thresholds. These results confirm the stable, superior performance of DE in smaller-sample settings.}

\begin{table}
    \centering
    \caption{{Simulation results for Setting 1 with $n=50$. Thresholds and achieved loss values are averaged over 100 repetitions, with standard errors in parentheses. Methods compared include proposed joint optimization with differential evolution (DE), greedy methods based on stepwise aggregation (SA) and stepwise splitting (SS), and principal amalgamation analysis (PAA). Oracle refers to the base thresholds $(70, 180, 250)$. Bold highlights the lowest loss values for each loss function.}}
    \label{tab: setting1_n50}
    \resizebox{\columnwidth}{!}{%
    \begin{tabular}{ccccccccccc}
    \toprule
      & & \multicolumn{4}{c}{$L_1$} & \multicolumn{4}{c}{$L_2$} & \multirow{2.4}{*}{PAA} \\
    \cmidrule(lr){3-6} \cmidrule(lr){7-10}
      Noise &  & Oracle & DE & SA & SS & Oracle & DE & SA & SS &  \\
    \midrule
        \multirow{4}{*}{$\nu=0$} & $t_1$ & 70 & 70.3 (0.0) & 70.0 (0.0) & 79.9 (0.6) & 70 & 70.8 (0.1) & 70.0 (0.0) & 70.6 (0.3) & 70.0 (0.0) \\
          & $t_2$ & 180 & 180.2 (0.1) & 180.1 (0.1) & 212.9 (2.1) & 180 & 179.8 (0.1) & 180.1 (0.1) & 181.4 (2.6) & 180.0 (0.0) \\
          & $t_3$ & 250 & 251.3 (0.1) & 250.2 (0.1) & 249.5 (0.6) & 250 & 252.6 (0.2) & 250.3 (0.1) & 262.0 (1.0) & 250.0 (0.0) \\
          & Loss  & 6.34 (.06) & \textbf{6.17 (.05)} & 6.35 (.06) & 62.0 (1.8) & 1.88 (.03) & \textbf{1.68 (.03)} & 1.93 (.03) & 12.4 (0.7) & -- \\
        \midrule
        \multirow{4}{*}{$\nu=5$} & $t_1$ & 70 & 73.3 (0.1) & 75.8 (0.4) & 81.2 (0.5) & 70 & 74.3 (0.2) & 74.7 (0.5) & 74.1 (0.7) & 74.4 (0.5) \\
          & $t_2$ & 180 & 180.1 (0.2) & 179.4 (0.4) & 218.1 (1.9) & 180 & 179.2 (0.2) & 178.9 (0.6) & 172.5 (2.8) & 178.4 (0.4) \\
          & $t_3$ & 250 & 255.4 (0.2) & 255.7 (0.5) & 255.0 (0.7) & 250 & 256.4 (0.2) & 257.1 (0.6) & 263.8 (1.2) & 255.7 (0.5) \\
          & Loss  & 17.7 (0.3) & \textbf{11.1 (0.1)} & 15.5 (0.4) & 70.3 (1.9) & 6.11 (.15) & \textbf{3.25 (.06)} & 5.18 (.16) & 15.9 (0.6) & -- \\
        \midrule
        \multirow{4}{*}{$\nu=10$} & $t_1$ & 70 & 73.8 (0.2) & 78.0 (0.7) & 83.0 (0.5) & 70 & 73.9 (0.3) & 77.9 (0.8) & 72.9 (0.8) & 65.7 (1.0) \\
          & $t_2$ & 180 & 178.8 (0.3) & 176.9 (0.9) & 220.5 (1.8) & 180 & 177.6 (0.4) & 176.5 (0.9) & 171.9 (3.0) & 160.1 (3.7) \\
          & $t_3$ & 250 & 258.3 (0.2) & 261.2 (0.9) & 260.1 (0.9) & 250 & 258.9 (0.4) & 263.7 (1.0) & 263.2 (1.4) & 241.7 (3.0) \\
          & Loss  & 33.6 (0.6) & \textbf{22.7 (0.4)} & 37.7 (1.1) & 91.7 (2.0) & 11.7 (0.3) & \textbf{7.35 (.18)} & 12.5 (0.4) & 22.0 (0.7) & -- \\
    \bottomrule
    \end{tabular}
    }
\end{table}

\begin{table}
    \centering
    \caption{{Simulation results for Setting 1 with $n=100$. Thresholds and achieved loss values are averaged over 100 repetitions, with standard errors in parentheses. Methods compared include proposed joint optimization with differential evolution (DE), greedy methods based on stepwise aggregation (SA) and stepwise splitting (SS), and principal amalgamation analysis (PAA). Oracle refers to the base thresholds $(70, 180, 250)$. Bold highlights the lowest loss values for each loss function.}}
    \label{tab: setting1_n100}
    \resizebox{\columnwidth}{!}{%
    \begin{tabular}{ccccccccccc}
    \toprule
      & & \multicolumn{4}{c}{$L_1$} & \multicolumn{4}{c}{$L_2$} & \multirow{2.4}{*}{PAA} \\
    \cmidrule(lr){3-6} \cmidrule(lr){7-10}
      Noise &  & Oracle & DE & SA & SS & Oracle & DE & SA & SS &  \\
    \midrule
        \multirow{4}{*}{$\nu=0$} & $t_1$ & 70 & 70.3 (0.0) & 70.0 (0.0) & 80.4 (0.5) & 70 & 70.6 (0.1) & 70.0 (0.0) & 70.0 (0.1) & 70.0 (0.0) \\
          & $t_2$ & 180 & 180.1 (0.1) & 180.1 (0.1) & 220.6 (1.7) & 180 & 179.9 (0.1) & 180.1 (0.1) & 186.2 (2.3) & 180.0 (0.0) \\
          & $t_3$ & 250 & 251.1 (0.1) & 250.0 (0.0) & 251.0 (0.4) & 250 & 252.8 (0.2) & 250.2 (0.1) & 262.0 (0.7) & 250.0 (0.0) \\
          & Loss  & 6.40 (.04) & \textbf{6.29 (.04)} & 6.41 (.04) & 66.4 (1.7) & 1.95 (.02) & \textbf{1.82 (.02)} & 1.96 (.02) & 11.1 (0.5) & -- \\
        \midrule
        \multirow{4}{*}{$\nu=5$} & $t_1$ & 70 & 73.5 (0.1) & 75.8 (0.4) & 81.6 (0.4) & 70 & 74.3 (0.1) & 75.6 (0.5) & 73.1 (0.4) & 73.9 (0.5) \\
          & $t_2$ & 180 & 179.9 (0.1) & 179.9 (0.4) & 222.9 (1.8) & 180 & 179.1 (0.2) & 180.0 (0.5) & 179.0 (2.7) & 178.5 (0.4) \\
          & $t_3$ & 250 & 255.3 (0.1) & 256.8 (0.5) & 255.9 (0.7) & 250 & 256.3 (0.2) & 257.1 (0.6) & 262.1 (0.6) & 255.0 (0.5) \\
          & Loss  & 17.1 (0.2) & \textbf{11.3 (0.1)} & 15.3 (0.4) & 78.8 (1.3) & 5.84 (.10) & \textbf{3.46 (.05)} & 5.14 (.15) & 14.5 (0.5) & -- \\
        \midrule
        \multirow{4}{*}{$\nu=10$} & $t_1$ & 70 & 73.7 (0.1) & 78.4 (0.7) & 82.6 (0.3) & 70 & 74.1 (0.2) & 77.6 (0.8) & 72.2 (0.3) & 66.9 (0.9) \\
          & $t_2$ & 180 & 178.5 (0.2) & 176.5 (1.0) & 227.9 (1.0) & 180 & 177.7 (0.3) & 177.2 (1.0) & 176.1 (3.0) & 170.7 (2.2) \\
          & $t_3$ & 250 & 258.1 (0.2) & 262.0 (0.9) & 263.7 (0.6) & 250 & 258.5 (0.3) & 261.7 (0.9) & 262.0 (1.0) & 250.5 (2.1) \\
          & Loss  & 33.6 (0.4) & \textbf{23.8 (0.2)} & 39.2 (1.1) & 97.2 (1.3) & 11.5 (0.2) & \textbf{8.00 (.12)} & 12.8 (0.3) & 22.0 (0.6) & -- \\
    \bottomrule
    \end{tabular}
    }
\end{table}

\begin{table}
    \centering
    \caption{{Simulation results for Setting 2 with $n=50$ empirical distributions and $K=3$ for $L_1$ loss and $K=2$ for $L_2$ loss and PAA. Thresholds and achieved loss values are averaged over 100 repetitions, with standard errors in parentheses. Oracle thresholds are $(70, 180, 250)$ for $L_1$ and $(70, 180)$ for $L_2$. Bold highlights the lowest loss values for each loss function.}}
    \label{tab: setting2_n50}
    \resizebox{\columnwidth}{!}{%
    \begin{tabular}{cccccccccccc}
    \toprule
      & & \multicolumn{4}{c}{$L_1$, $K=3$} & \multicolumn{4}{c}{$L_2$, $K=2$} & \multirow{2.4}{*}{PAA, $K=2$}\\
    \cmidrule(lr){3-6} \cmidrule(lr){7-10}
      Noise &  & Oracle & DE & SA & SS & Oracle & DE & SA & SS & \\
    \midrule
        \multirow{4}{*}{$\nu=0$} & $t_1$ & 70 & 70.2 (0.0) & 70.0 (0.0) & 68.3 (0.1) & 70 & 77.3 (0.7) & 70.0 (0.0) & 129.4 (0.8) & 70.0 (0.0) \\
          & $t_2$ & 180 & 179.9 (0.1) & 180.2 (0.1) & 226.8 (0.6) & 180 & 197.2 (2.1) & 179.1 (1.2) & 313.5 (13.6) & 183.8 (1.2) \\
          & $t_3$ & 250 & 250.7 (0.1) & 250.1 (0.0) & 252.3 (0.2) & -- & -- & -- & -- & -- \\
          & Loss  & 6.59 (.06) & \textbf{6.42 (.06)} & 6.62 (.06) & 39.4 (0.7) & 3.03 (.05) & \textbf{2.54 (.04)} & 4.27 (.41) & 10.7 (0.1) & -- \\
        \midrule
        \multirow{4}{*}{$\nu=5$} & $t_1$ & 70 & 74.4 (0.2) & 77.5 (0.4) & 74.4 (0.2) & 70 & 78.8 (0.4) & 77.0 (0.4) & 129.2 (0.2) & 63.7 (0.4) \\
          & $t_2$ & 180 & 178.4 (0.2) & 179.4 (0.5) & 226.0 (1.0) & 180 & 201.8 (1.0) & 178.8 (1.4) & 274.4 (15.6) & 84.1 (2.5) \\
          & $t_3$ & 250 & 253.6 (0.1) & 254.6 (0.4) & 256.2 (0.4) & -- & -- & -- & -- & -- \\
          & Loss  & 19.9 (0.4) & \textbf{12.2 (0.2)} & 17.2 (0.5) & 48.2 (0.9) & 10.6 (0.3) & \textbf{4.26 (.06)} & 8.49 (.35) & 14.4 (0.2) & -- \\
        \midrule
        \multirow{4}{*}{$\nu=10$} & $t_1$ & 70 & 76.2 (0.2) & 80.2 (0.7) & 77.1 (0.2) & 70 & 79.9 (0.3) & 80.0 (0.7) & 127.3 (0.2) & 56.4 (0.6) \\
          & $t_2$ & 180 & 175.3 (0.3) & 173.9 (0.9) & 225.1 (1.1) & 180 & 203.8 (0.7) & 175.6 (1.8) & 162.8 (15.5) & 80.9 (0.6) \\
          & $t_3$ & 250 & 255.4 (0.2) & 258.0 (0.8) & 261.2 (0.4) & -- & -- & -- & -- & -- \\
          & Loss  & 38.9 (0.7) & \textbf{25.4 (0.4)} & 40.8 (1.1) & 65.4 (1.2) & 20.0 (0.4) & \textbf{9.61 (.18)} & 19.9 (0.5) & 26.8 (0.4) & -- \\
    \bottomrule
    \end{tabular}
    }
\end{table}

\begin{table}
    \centering
    \caption{{Simulation results for Setting 2 with $n=100$ empirical distributions and $K=3$ for $L_1$ loss and $K=2$ for $L_2$ loss and PAA. Thresholds and achieved loss values are averaged over 100 repetitions, with standard errors in parentheses. Oracle thresholds are $(70, 180, 250)$ for $L_1$ and $(70, 180)$ for $L_2$. Bold highlights the lowest loss values for each loss function.}}
    \label{tab: setting2_n100}
    \resizebox{\columnwidth}{!}{%
    {
    \begin{tabular}{cccccccccccc}
    \toprule
      & & \multicolumn{4}{c}{$L_1$, $K=3$} & \multicolumn{4}{c}{$L_2$, $K=2$} & \multirow{2.4}{*}{PAA, $K=2$}\\
    \cmidrule(lr){3-6} \cmidrule(lr){7-10}
      Noise &  & Oracle & DE & SA & SS & Oracle & DE & SA & SS & \\
    \midrule
        \multirow{4}{*}{$\nu=0$} & $t_1$ & 70 & 70.2 (0.0) & 70.0 (0.0) & 68.2 (0.1) & 70 & 75.6 (0.5) & 70.0 (0.0) & 130.3 (0.1) & 70.0 (0.0) \\
          & $t_2$ & 180 & 179.7 (0.1) & 180.0 (0.1) & 227.4 (0.4) & 180 & 192.4 (1.5) & 179.1 (1.0) & 334.3 (12.5) & 180.5 (0.3) \\
          & $t_3$ & 250 & 250.8 (0.1) & 250.0 (0.0) & 252.3 (0.2) & -- & -- & -- & -- & -- \\
          & Loss  & 6.54 (.04) & \textbf{6.42 (.04)} & 6.57 (.04) & 39.6 (0.5) & 2.94 (.03) & \textbf{2.61 (.02)} & 3.85 (.27) & 10.6 (0.1) & -- \\
        \midrule
        \multirow{4}{*}{$\nu=5$} & $t_1$ & 70 & 74.5 (0.1) & 77.7 (0.4) & 74.5 (0.2) & 70 & 78.6 (0.3) & 77.8 (0.4) & 129.3 (0.1) & 63.6 (0.3) \\
          & $t_2$ & 180 & 178.4 (0.1) & 179.2 (0.4) & 227.0 (0.4) & 180 & 200.8 (0.6) & 179.8 (0.9) & 291.6 (14.9) & 78.4 (0.3) \\
          & $t_3$ & 250 & 253.9 (0.1) & 254.9 (0.4) & 256.7 (0.1) & -- & -- & -- & -- & -- \\
          & Loss  & 20.3 (0.3) & \textbf{12.5 (0.1)} & 17.9 (0.4) & 47.4 (0.6) & 10.6 (0.2) & \textbf{4.50 (.04)} & 8.18 (.25) & 14.8 (0.2) & -- \\
        \midrule
        \multirow{4}{*}{$\nu=10$} & $t_1$ & 70 & 76.0 (0.1) & 79.9 (0.8) & 77.2 (0.2) & 70 & 80.4 (0.3) & 80.3 (0.8) & 127.3 (0.2) & 56.9 (0.5) \\
          & $t_2$ & 180 & 175.1 (0.3) & 175.6 (1.0) & 225.8 (0.6) & 180 & 206.2 (0.8) & 172.8 (1.8) & 146.7 (15.0) & 82.1 (0.6) \\
          & $t_3$ & 250 & 255.1 (0.1) & 260.5 (0.7) & 260.9 (0.3) & -- & -- & -- & -- & -- \\
          & Loss  & 39.0 (0.5) & \textbf{26.8 (0.3)} & 41.6 (1.2) & 65.0 (0.7) & 20.1 (0.3) & \textbf{10.5 (0.1)} & 20.6 (0.5) & 28.0 (0.3) & -- \\
    \bottomrule
    \end{tabular}
    }
    }
\end{table}

\section{Additional results for real data experiments}\label{sec: supplement-real}

\subsection{{Screeplots for real data experiments}}

{We report screeplots of the optimized empirical loss as the number of thresholds varies for the four CGM datasets presented in the main text: the healthy cohort from \citet{shahContinuousGlucoseMonitoring2019}, the type 1 diabetes (T1D) cohort from \citet{brownSixMonthRandomizedMulticenter2019}, the combined healthy + T1D analysis, and the AI-READI analysis.     
Consistent with the main text, the separate healthy and T1D analyses use the $L_1$ loss, whereas the combined and AI-READI analyses use the $L_2$ loss. Plots presented in Figure~S4 suggest that, across all datasets, $K=4$ thresholds provide an optimal balance between simplicity and distributional preservation.}

\begin{figure}[t]
    \centering
    \includegraphics[width=0.48\linewidth]{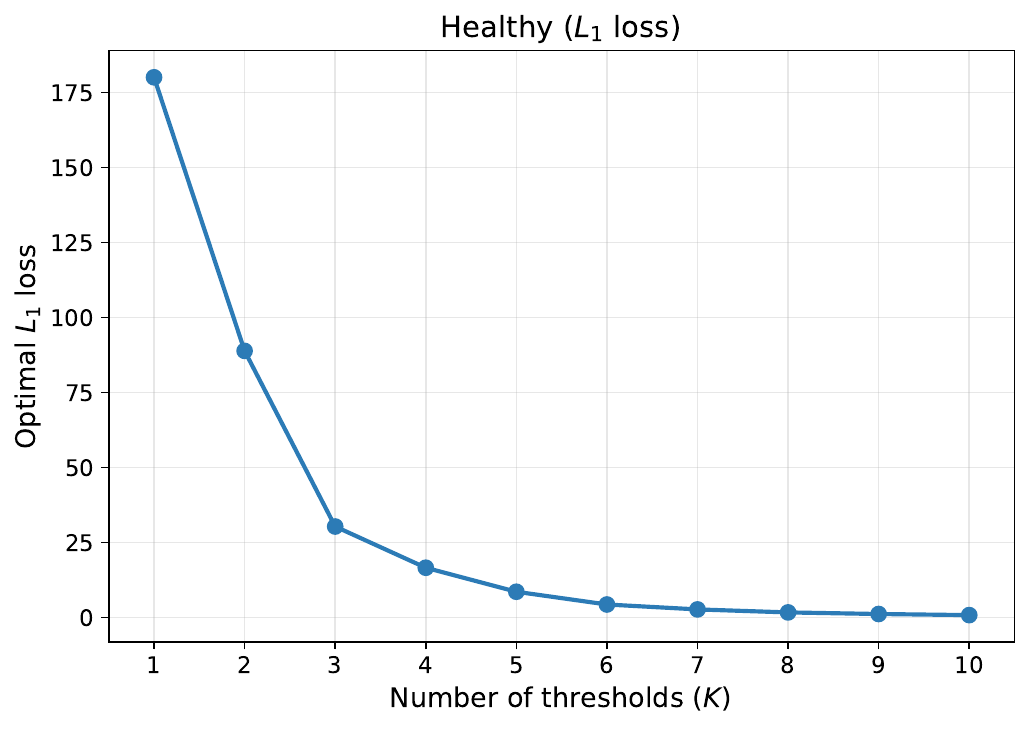}
    \hfill
    \includegraphics[width=0.48\linewidth]{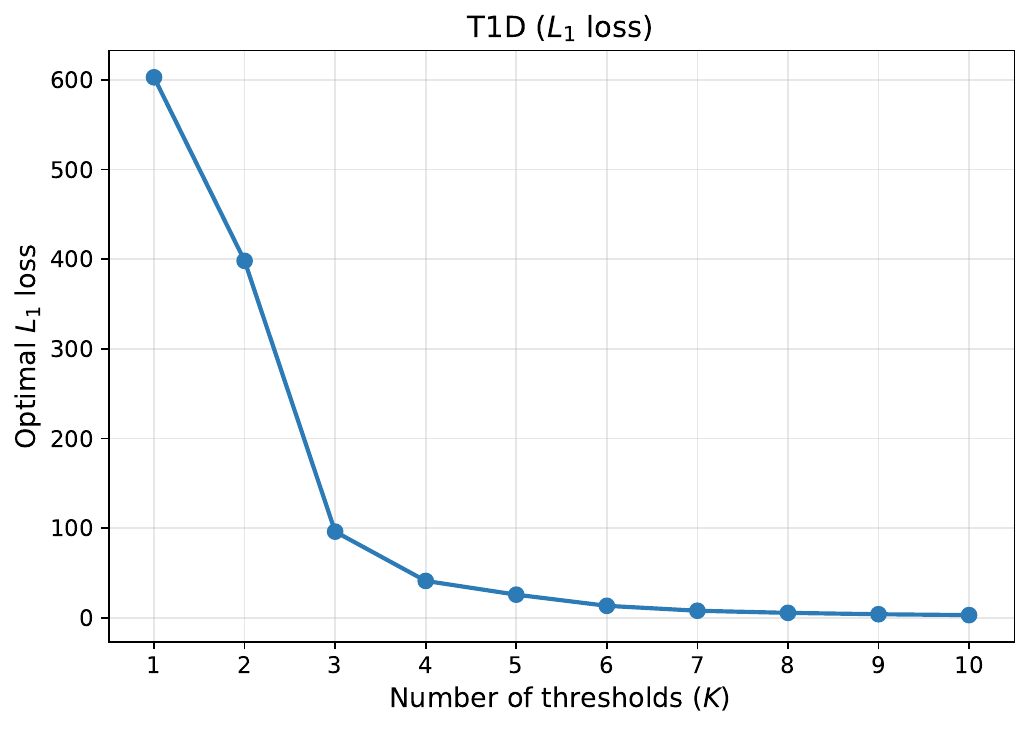}

    \vspace{0.5em}

    \includegraphics[width=0.48\linewidth]{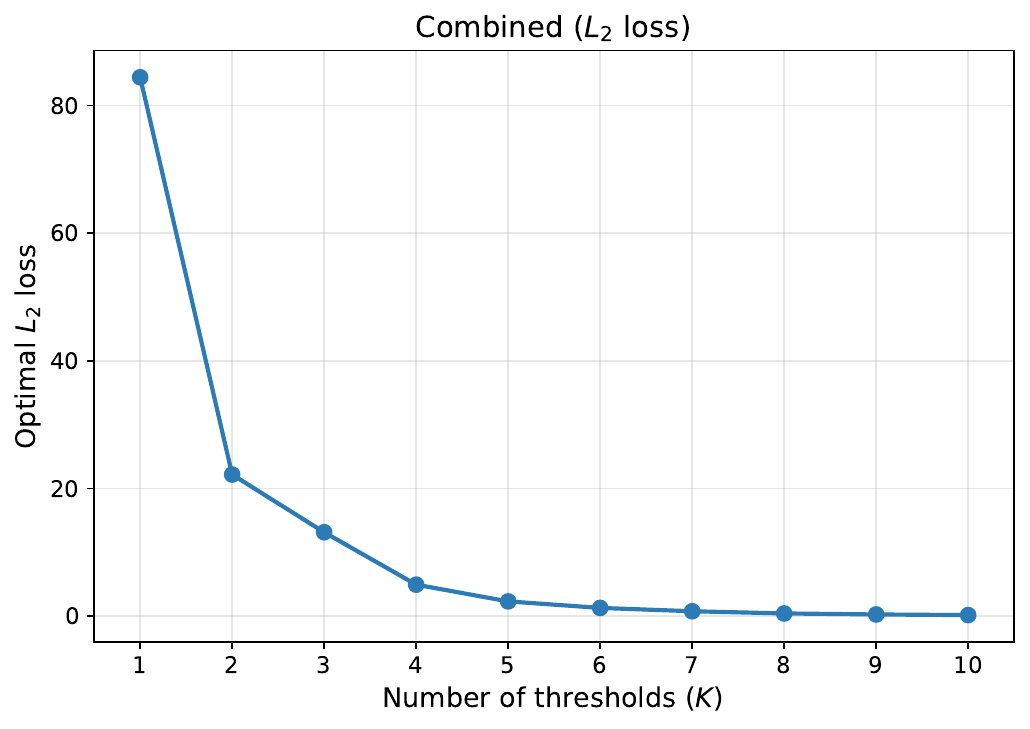}
    \hfill
    \includegraphics[width=0.48\linewidth]{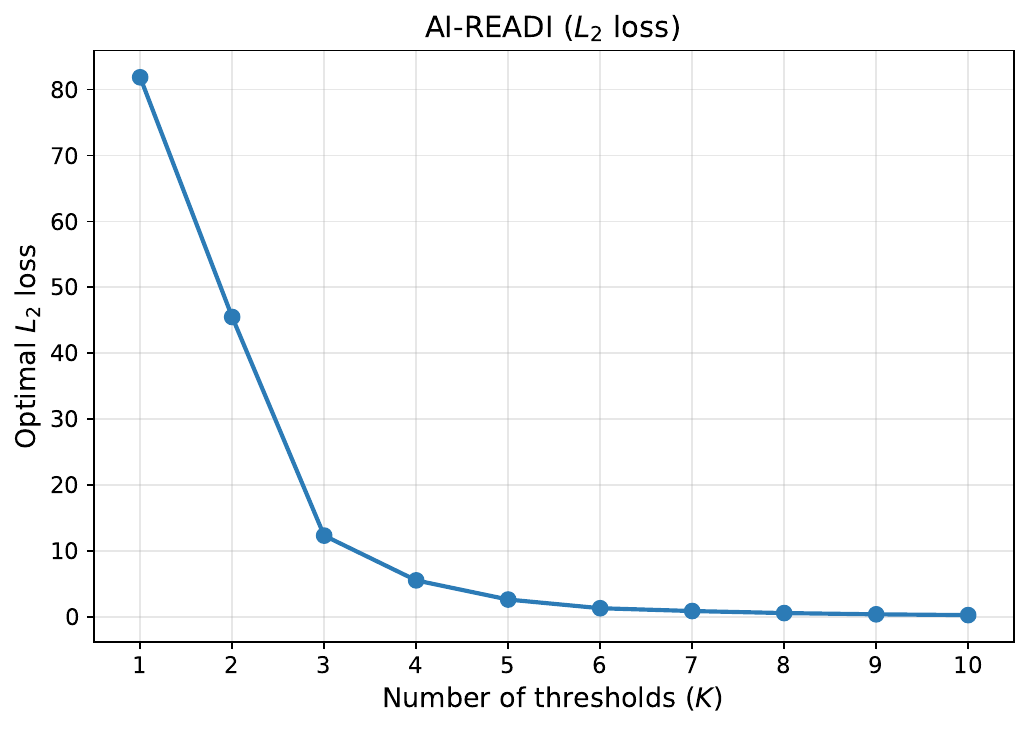}
    \caption{{Screeplots for the four real-data experiments in the main manuscript. The panels are ordered as follows: non-diabetes data from \citet{shahContinuousGlucoseMonitoring2019} (top left), type 1 diabetes data from \citet{brownSixMonthRandomizedMulticenter2019} (top right), the combined non- and type 1 diabetes data (bottom left), and the AI-READI data (bottom right). The top panels use the $L_1$ loss, while the bottom panels use the $L_2$ loss.}}
    \label{fig:real-screeplots}
\end{figure}

\subsection{Additional separate analysis}

\subsubsection*{$L_2$ loss results}

We report the $L_2$ loss thresholds obtained by DE for the datasets of individuals without diabetes \citep{shahContinuousGlucoseMonitoring2019} and those with type 1 diabetes \citep{brownSixMonthRandomizedMulticenter2019}. 

For the data from individuals without diabetes \citep{shahContinuousGlucoseMonitoring2019}, DE with $K=4$ yields thresholds at $\{75, 102,  128, 164\}$ mg/dL with $L_2 = 1.15$, a 97\% reduction from $L_2 = 43.7$ at the four consensus thresholds $\{54, 70, 181, 251\}$ mg/dL. The $K=2$ data-driven thresholds are obtained at $\{120, 164\}$ mg/dL with $L_2=16.1$, still outperforming the four consensus thresholds. Although different from the $L_1$-derived thresholds, these $L_2$-derived thresholds again suggest capturing narrower ranges for more informative summaries of CGM distributions for individuals without diabetes. Focusing on the $K=2$ case, we also extract an interesting insight: while $L_1$-derived thresholds $\{72,128\}$ mg/dL generally capture the distributional structures, the higher $L_2$-derived thresholds $\{124, 164\}$ mg/dL highlight that differences in glucose distributions among individuals without diabetes typically arise in the higher glucose ranges.

For the type 1 diabetes data \citep{brownSixMonthRandomizedMulticenter2019}, DE with $K=4$ obtains thresholds at $\{92, 173, 270, 400\}$ mg/dL, which is more inflated than the $L_1$-derived thresholds. These thresholds yield $L_2 = 1.69$, a 86\% reduction from $L_2 = 12.2$ obtained at the consensus thresholds. The $K=2$ thresholds found by DE are $\{183, 281\}$ mg/dL with $L_2 = 9.47$, which still improves the $L_2$ loss over the four consensus thresholds by 22\%. These elevated $L_2$-driven thresholds indicate that differences in glucose patterns among individuals with type 1 diabetes often arise in the hyperglycemic ranges. Notably, the threshold at 400 mg/dL indicates a separation between individuals who hit the measurement limit and those whose glucose levels remain within the standard range of CGM devices.

\subsubsection*{Quantile plot for type 1 diabetes dataset}

\begin{figure}[htbp]
    \centering
    \includegraphics[width=\linewidth]{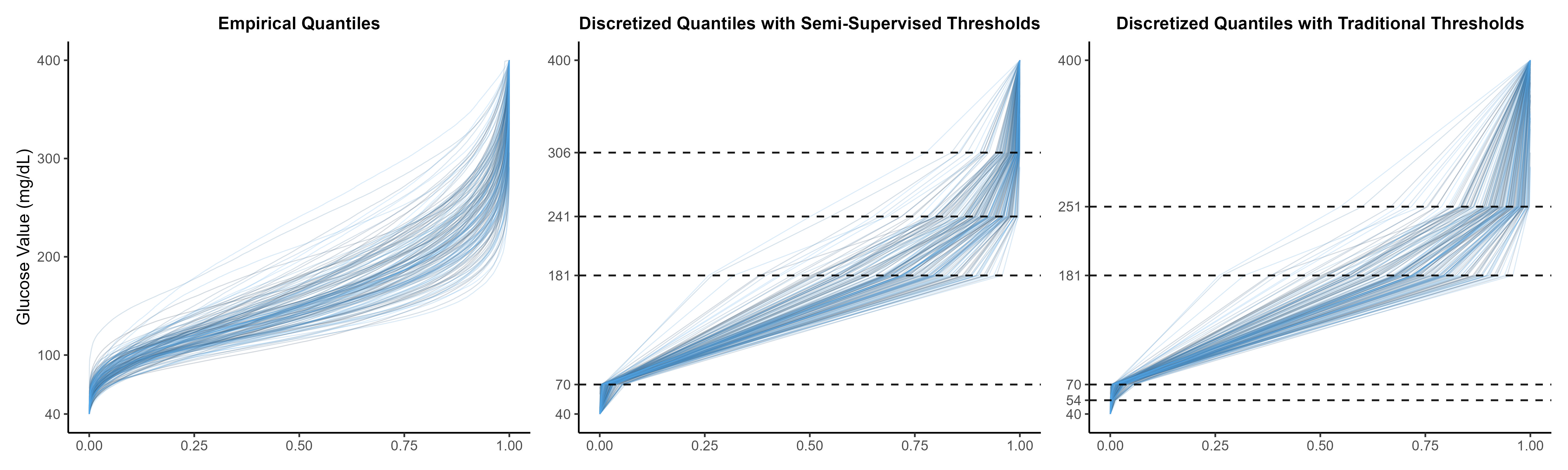}
    \caption{Quantile plots of glucose values for $n_2 = 168$ individuals with type 1 diabetes from \citet{brownSixMonthRandomizedMulticenter2019}, where each line represents a single individual. Empirical quantiles are shown in the left panel, while the middle and right panels display quantiles discretized using semi-supervised thresholds ($L_1$ loss, $K = 4$, fixed 70 \& 181 mg/dL) and traditional fixed thresholds, respectively, with thresholds indicated by horizontal lines. } 
    \label{fig: brown_quantiles}
\end{figure}

We also provide visualizations for the dataset of individuals with type 1 diabetes \citep{brownSixMonthRandomizedMulticenter2019}. To further illustrate the effectiveness of the semi-supervised thresholds, we depict empirical quantiles of individuals with type 1 diabetes and their piecewise linearizations using the semi-supervised (obtained in the real data experiment) and the consensus thresholds in \cref{fig: brown_quantiles}. It is observed that the semi-supervised data-driven thresholds better capture the distributional structures in the higher glycemic region than the consensus thresholds while preserving the standard thresholds $\{70, 181\}$ mg/dL for the standard TIR metric.

\subsection{Additional results on combined data}

\subsubsection*{$K=4$ with $L_2$ loss}
\begin{figure}[H]
    \centering
    \includegraphics[width=0.85\linewidth]{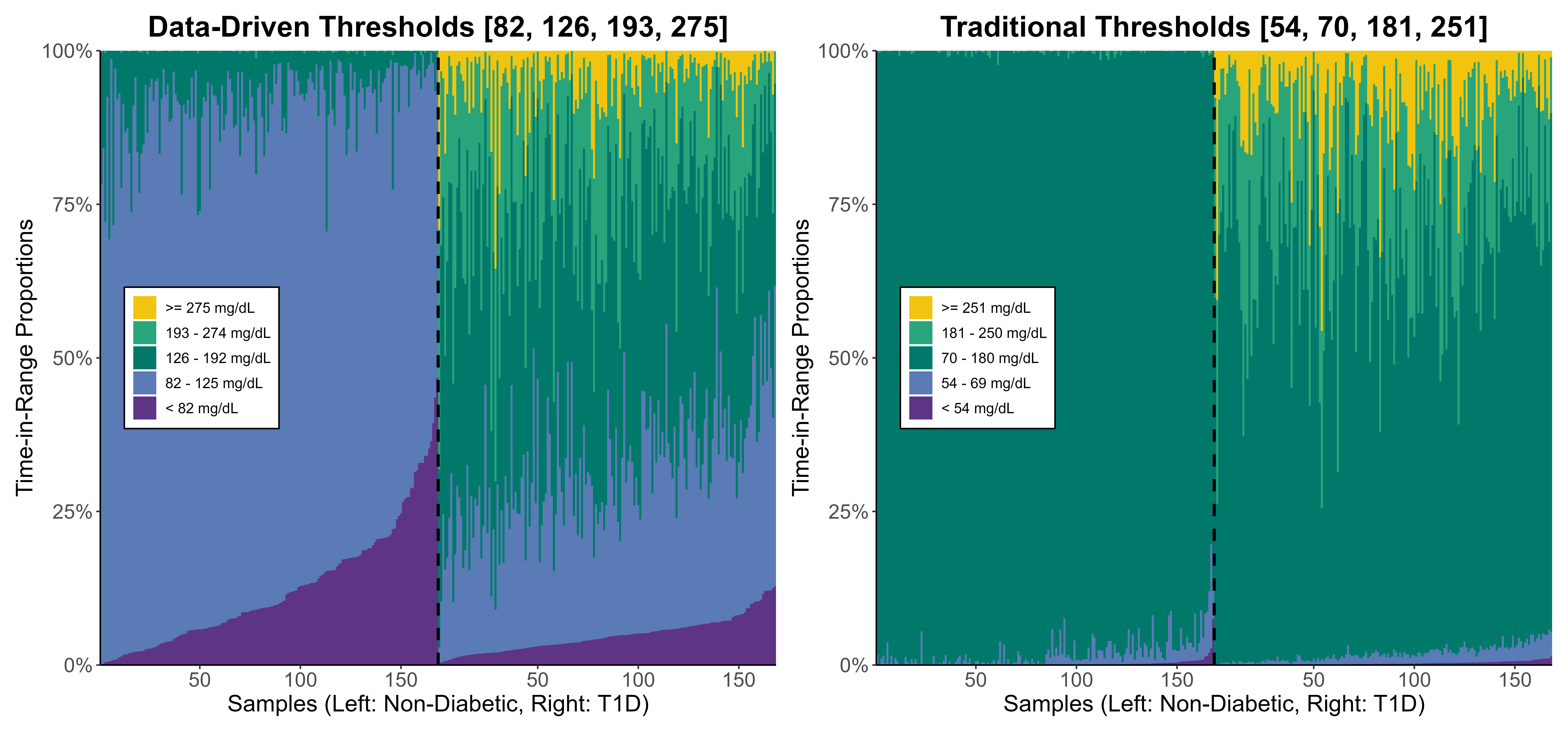}
    \caption{Compositional bar plots showing TIR proportions derived by data-driven thresholds ($L_2$ loss, $K = 4$; left) and traditional thresholds (right). Individuals without diabetes and those with type 1 diabetes are separated by vertical black lines, with samples arranged by time below range within each group.}
    \label{fig: k4 TIR barplots}
\end{figure}

We also present experimental results with $K=4$ thresholds. DE with $L_2$ loss yields thresholds at $\{82, 126, 193, 275\}$ mg/dL, which improves the $L_2$ loss by 97\% from that achieved using the four consensus thresholds. \cref{fig: k4 TIR barplots} illustrates the compositional barplots using the data-driven and consensus thresholds. As with the case $K=2$, the data-driven thresholds reveal clearly distinguishable differences through all TIR proportions defined by them. In contrast, the first two TIR ranges defined by traditional thresholds, below 54 mg/dL and 54--69 mg/dL, fail to capture differences between individuals without diabetes and those with type 1 diabetes. \cref{tab: combined_data_k4} reports the logistic regression results similarly to the $K=2$ case, showing that our data-driven thresholds still yield almost perfect classification with induced TIR proportions, except for TIR $<82$ mg/dL in which a moderate separation is achieved.

\begin{table}[t]
\centering
\caption{Univariate logistic regression results for each TIR proportion using data-driven (DE) thresholds $\{82, 126, 193, 275\}$ mg/dL versus consensus thresholds $\{54,70,181,251\}$ mg/dL.}
\label{tab: combined_data_k4}
\resizebox{\columnwidth}{!}{%
\begin{tabular}{lcclcc}
\toprule
\multicolumn{3}{c}{DE} & \multicolumn{3}{c}{Consensus} \\
\cmidrule(lr){1-3}\cmidrule(lr){4-6}
Ranges (mg/dL) & Decision Boundary & Accuracy (\%) & Ranges (mg/dL) & Decision Boundary & Accuracy (\%) \\
\midrule
TIR $<83$ & 0.074 & 74.4 & TIR $<54$ & 0.003 & 60.1 \\
TIR 82--125 & 0.522 & 99.4  & TIR 54--69 & 0.017 & 47.9 \\
TIR 126--192 & 0.266 & 97.9 & TIR 70--180 & 0.893 & 97.9 \\
TIR 193--274 & 0.021 & 100.0  & TIR 181--250 & 0.032 & 100.0 \\
TIR $\ge 275$ & 0.001 & 98.8  & TIR $\ge 251$ & 0.003 & 99.4 \\
\bottomrule
\end{tabular}
}
\end{table}

\subsubsection*{$L_1$ loss results}

Thresholds found by DE with the $L_1$ loss are presented for completeness. With $K=2$, DE identifies thresholds at $\{133, 247\}$ mg/dL, which yields similar classification results to the $L_2$ loss results. The $K=4$ thresholds are found at $\{77, 124, 195, 280\}$ mg/dL. Similar logistic regression accuracies based on the five TIR summaries from these $L_1$ thresholds are 61.9\%, 100.0\%, 97.0\%, 100.0\%, and 98.5\%, in increasing order of the corresponding glucose ranges. The lowest range $<77$ mg/dL performs considerably worse than TIR $<82$ mg/dL from the $L_2$ loss result, emphasizing that small change in glucose thresholds can result in different downstream statistical power. Given that TIR $<70$ mg/dL fails to discriminate the two groups as in Table 3, the lower glycemic region around 70-80 appears highly sensitive to threshold choice in discriminating the two groups. Thus, we interpret the $L_1$-derived thresholds as emphasizing reconstruction of distributions rather than discrimination, whereas the $L_2$-derived thresholds adapt to this sensitive region.


\subsection{{Additional comparison with naive thresholds}}

{
We additionally compare the performance of Naive thresholds, which are defined as tertiles ($K=2$) and quintiles ($K=4$) of the pooled glucose measurements across all subjects.
Table~\ref{tab:real_all_thresholds} reports the consensus, DE, and Naive thresholds. The Naive thresholds are considerably narrower than both consensus and DE thresholds, reflecting the concentration of the pooled glucose distribution.

\begin{table}[h]
\centering 
\caption{ Thresholds (mg/dL) used for each method and dataset in the real CGM data experiments. Consensus denotes standard clinical cutoffs, DE denotes data-driven thresholds obtained by differential evolution, and Naive denotes pooled-quantile thresholds from the empirical glucose distribution. For DE, the Healthy \citep{shahContinuousGlucoseMonitoring2019} and type 1 diabetes (T1D) \citep{brownSixMonthRandomizedMulticenter2019} datasets used $L_1$ loss, whereas the Combined and AI-READI datasets used $L_2$ loss.}
\label{tab:real_all_thresholds}
\footnotesize
\begin{tabular}{lcccccc}
\toprule
 & \multicolumn{3}{c}{$K=2$} & \multicolumn{3}{c}{$K=4$} \\
\cmidrule(lr){2-4} \cmidrule(lr){5-7}
Dataset & Consensus & DE & Naive & Consensus & DE & Naive \\
\midrule
Healthy & \multirow{4}{*}{$70, 181$} & $72, 128$ & $91, 104$ & \multirow{4}{*}{$54, 70, 181, 251$} & $76, 101, 124, 155$ & $85, 93, 101, 112$ \\
T1D & & $211, 289$ & $126, 176$ & & $85, 172, 233, 302$ & $110, 135, 164, 207$ \\
Combined & & $150, 258$ & $123, 173$ & & $82, 126, 193, 275$ & $107, 132, 161, 204$ \\
AI-READI & & $96, 170$ & $109, 129$ & & $90, 128, 172, 232$ & $102, 113, 124, 141$ \\
\bottomrule
\end{tabular}
\end{table}

We examine the same three tasks as in Section~5: preservation of the underlying distributional information, discrimination on the combined cohort, and downstream prediction in AI-READI. 

\subsubsection*{Loss minimization}
The loss ($L_1$ and $L_2$) comparisons quantify how much information is lost when the full glucose distributions are compressed into threshold-based summaries.
Table~\ref{tab:real_loss_comparison} compares losses using the same criterion emphasized in the main real-data analyses: $L_1$ for the separate healthy and type 1 diabetes cohorts, and $L_2$ for the combined and AI-READI cohorts. 
DE consistently attains the smallest loss values, indicating that it preserves the underlying distributional information more effectively than either consensus or Naive thresholds. Naive thresholds perform particularly poorly for the distance-preserving $L_2$ loss and are also inferior to the consensus thresholds under $L_1$ in the type 1 diabetes dataset. An exception is the non-diabetes (Healthy) dataset under $L_1$, where the glucose distributions are most concentrated and therefore better aligned with narrow Naive thresholds. Overall, although Naive thresholds are also data-driven, they do not preserve distributional shapes or between-subject distances as well as DE.

\begin{table}[h]
\centering
\caption{Comparison of loss functions across consensus, DE, and pooled-quantile naive thresholds for the CGM datasets. Healthy and T1D rows report $L_1$ loss, whereas Combined and AI-READI rows report $L_2$ loss. The smallest loss value in each comparable $K$ block is shown in bold.}
\label{tab:real_loss_comparison}
\small
\begin{tabular}{llrrrrrr}
\toprule
 & & \multicolumn{3}{c}{$K=2$} & \multicolumn{3}{c}{$K=4$} \\
\cmidrule(lr){3-5} \cmidrule(lr){6-8}
Dataset & Loss & Consensus & DE & Naive & Consensus & DE & Naive \\
\midrule
Healthy & $L_1$ & 656.9 & \textbf{88.9} & 371.4 & 655.9 & \textbf{16.6} & 183.3 \\
\midrule
T1D & $L_1$ & 1236.0 & \textbf{398.2} & 1518.7 & 159.9 & \textbf{41.2} & 615.7 \\
\midrule
Combined & $L_2$ & 111.6 & \textbf{22.2} & 577.3 & 144.2 & \textbf{4.9} & 244.7 \\
\midrule
AI-READI & $L_2$ & 103.4 & \textbf{45.5} & 454.2 & 121.1 & \textbf{5.5} & 289.0 \\
\bottomrule
\end{tabular}
\end{table}

\subsubsection*{Discriminative performance on the combined cohort}

We next add Naive thresholds to the combined-data discrimination analysis from Section~5.2. Tables~\ref{tab:combined_logreg_k2} and \ref{tab:combined_logreg_k4} report the univariate logistic-regression accuracy of each TIR component for classifying individuals without diabetes versus those with type 1 diabetes. Since each method defines different TIR ranges, we compare methods by their average and minimum accuracy across all bins. 
Naive thresholds improve on the consensus thresholds overall but underperform DE. In particular, for $K=4$, Naive thresholds produce one bin with accuracy only 55.7\%, whereas DE yields more consistent and superior average discriminative performance across all bins.

\begin{table}[h]
\centering
\caption{Univariate logistic-regression accuracy (\%) on the combined cohort using $K=2$ thresholds ($L_2$ loss for DE). Each TIR proportion is used as the sole predictor for classifying individuals without diabetes vs.\ with type 1 diabetes.}
\label{tab:combined_logreg_k2}
\small
\begin{tabular}{lclclc}
\toprule
\multicolumn{2}{c}{Consensus} & \multicolumn{2}{c}{DE} & \multicolumn{2}{c}{Naive} \\
\cmidrule(lr){1-2} \cmidrule(lr){3-4} \cmidrule(lr){5-6}
Range & Accuracy(\%) & Range & Accuracy(\%) & Range & Accuracy(\%) \\
\midrule
TIR $<70$ & 47.0 & TIR $<150$ & 100.0 & TIR $<123$ & 100.0 \\
TIR 70--180 & 97.9 & TIR 150--257 & 100.0 & TIR 123--172 & 94.3 \\
TIR $\geq 181$ & 100.0 & TIR $\geq 258$ & 99.1 & TIR $\geq 173$ & 100.0 \\
\bottomrule
\end{tabular}
\end{table}

\begin{table}[h]
\centering
\caption{Univariate logistic-regression accuracy (\%) on the combined cohort using $K=4$ thresholds ($L_2$ loss for DE). Each TIR proportion is used as the sole predictor for classifying individuals without diabetes vs.\ with type 1 diabetes.}
\label{tab:combined_logreg_k4}
\small
\begin{tabular}{lclclc}
\toprule
\multicolumn{2}{c}{Consensus} & \multicolumn{2}{c}{DE} & \multicolumn{2}{c}{Naive} \\
\cmidrule(lr){1-2} \cmidrule(lr){3-4} \cmidrule(lr){5-6}
Range & Accuracy(\%) & Range & Accuracy(\%) & Range & Accuracy(\%) \\
\midrule
TIR $<54$ & 60.1 & TIR $<82$ & 74.4 & TIR $<107$ & 98.2 \\
TIR 54--69 & 47.9 & TIR 82--125 & 99.4 & TIR 107--131 & 55.7 \\
TIR 70--180 & 97.9 & TIR 126--192 & 97.9 & TIR 132--160 & 94.9 \\
TIR 181--250 & 100.0 & TIR 193--274 & 100.0 & TIR 161--203 & 100.0 \\
TIR $\geq 251$ & 99.4 & TIR $\geq 275$ & 98.8 & TIR $\geq 204$ & 100.0 \\
\bottomrule
\end{tabular}
\end{table}

\subsubsection*{Downstream prediction in AI-READI}

Finally, we compare the performance of Naive thresholds in the AI-READI linear-model analysis. Table~\ref{tab:full_lm_results} shows that Naive thresholds remain inferior to WR overall, but interestingly, they perform competitively with and sometimes significantly better than DE for lipid outcomes ($K=2$ for HDL-C and TG/HDL-C; $K=4$ for TG/HDL-C) according to Clarke's test for non-nested models. This pattern may indicate that glucose--lipid associations in this cohort are largely driven by the narrow, moderate glucose ranges around 100--140 mg/dL, where the Naive thresholds concentrate.

To assess whether this advantage generalizes, Table~\ref{tab:full_lm_results_hba1c} repeats the comparison for HbA1c. Here the Naive thresholds lose their advantage: they perform significantly worse than DE and are even inferior to consensus thresholds for $K=2$. DE achieves the best performance among threshold-based predictions and again closely tracks WR when $K=4$. Taken together, the lipid and HbA1c results suggest that Naive thresholds can be competitive when the signal is concentrated in a narrow middle glucose region, but this does not generalize to HbA1c, where DE remains consistently strong.

\begin{table}[h]
\centering
\caption{Comparison of model fits for lipid variables with TIR compositional predictors based on consensus (CS), data-driven (DE), Naive pooled-quantile thresholds, and Wasserstein regression (WR) with full distributions as predictors. $\Delta$AIC denotes the difference in AIC relative to DE (AIC$_{\text{CS}}$ $-$ AIC$_{\text{DE}}$ and AIC$_{\text{Naive}}$ $-$ AIC$_{\text{DE}}$), whose larger value indicates DE achieves better fit. Significant $p$-values from Clarke's test for non-nested model comparison are indicated with $\Delta$AIC.}
\label{tab:full_lm_results}
\small
\begin{tabular}{llrrr}
\toprule
$K$ & Metric & \multicolumn{1}{c}{HDL-C} & \multicolumn{1}{c}{TG} & \multicolumn{1}{c}{TG/HDL-C} \\
\midrule
\multirow{5}{*}{$K=2$} & $R^2_{\text{CS}}$ & 0.011 & 0.025 & 0.026 \\
 & $R^2_{\text{DE}}$ & 0.018 & 0.049 & 0.050 \\
 & $R^2_{\text{Naive}}$ & 0.035 & 0.050 & 0.059 \\
 & $\Delta$AIC (CS$-$DE) & 4.2$^\ddagger$ & 14.5$^\ddagger$ & 14.3$^\ddagger$ \\
 & $\Delta$AIC (Naive$-$DE) & -10.3$^\ddagger$ & -0.8 & -5.5$^\dagger$ \\
\midrule
\multirow{5}{*}{$K=4$} & $R^2_{\text{CS}}$ & 0.014 & 0.040 & 0.040 \\
 & $R^2_{\text{DE}}$ & 0.041 & 0.052 & 0.060 \\
 & $R^2_{\text{Naive}}$ & 0.040 & 0.053 & 0.062 \\
 & $\Delta$AIC (CS$-$DE) & 16.2$^\ddagger$ & 7.5$^*$ & 11.8$^\dagger$ \\
 & $\Delta$AIC (Naive$-$DE) & 0.9 & -0.3 & -1.3$^\dagger$ \\
\midrule
Full distribution & $R^2_{\text{WR}}$ & 0.041 & 0.054 & 0.067 \\
\bottomrule
\end{tabular}

\vspace{0.5mm}
{\footnotesize $^*$ $p<0.05$, $^\dagger$ $p<0.01$, $^\ddagger$ $p<0.001$.\hspace{19em}}
\end{table}

\begin{table}[h]
\centering
\caption{Comparison of model fits for HbA1c with TIR compositional predictors based on consensus (CS), data-driven (DE), Naive pooled-quantile thresholds, and Wasserstein regression (WR) with full distributions as predictors. $\Delta$AIC denotes the difference in AIC relative to DE (AIC$_{\text{CS}}$ $-$ AIC$_{\text{DE}}$ and AIC$_{\text{Naive}}$ $-$ AIC$_{\text{DE}}$), whose larger values indicate DE achieves better fit. Significant $p$-values from Clarke's test for non-nested model comparison are indicated with $\Delta$AIC.}
\label{tab:full_lm_results_hba1c}
\vskip 3mm
\small
\begin{tabular}{llr}
\toprule
$K$ & Metric & \multicolumn{1}{c}{HbA1c} \\
\midrule
\multirow{5}{*}{$K=2$} & $R^2_{\text{CS}}$ & 0.217 \\
 & $R^2_{\text{DE}}$ & 0.225 \\
 & $R^2_{\text{Naive}}$ & 0.208 \\
 & $\Delta$AIC (CS$-$DE) & 5.6$^*$ \\
 & $\Delta$AIC (Naive$-$DE) & 12.5$^*$ \\
\midrule
\multirow{5}{*}{$K=4$} & $R^2_{\text{CS}}$ & 0.227 \\
 & $R^2_{\text{DE}}$ & 0.248 \\
 & $R^2_{\text{Naive}}$ & 0.231 \\
 & $\Delta$AIC (CS$-$DE) & 15.9$^\ddagger$ \\
 & $\Delta$AIC (Naive$-$DE) & 13.4$^\ddagger$ \\
\midrule
Full distribution & $R^2_{\text{WR}}$ & 0.255 \\
\bottomrule
\end{tabular}

\vspace{0.5mm}
{\footnotesize $^*$ $p<0.05$, $^\dagger$ $p<0.01$, $^\ddagger$ $p<0.001$.\hspace{8em}}
\end{table}

}

\section{Differential evolution algorithm}\label{sec:supp-DE}
To implement differential evolution, we employ the \texttt{differential\_evolution} function from the Python library \texttt{SciPy} \citep{virtanenSciPy10Fundamental2020}. The key default hyperparameters are: population size $P = 15K$, crossover probability $C_r=0.7$, and the mutation factors $F_m$, uniformly sampled from $(0.5, 1)$ for each mutation. The initial population is sampled via Latin Hypercube sampling, which aims to maximize coverage of the available parameter space.
\cref{alg:de-thresholds} details the DE algorithm implemented in our experiments. In the semisupervised case with fixed thresholds $\bt_{\text{fix}}$, we run the same algorithm as below but incorporate $\bt_{\text{fix}}$ in the loss evaluation by sorting thresholds $\bt' = \bt\cup\bt_{\text{fix}}$ and computing $L(\bt';\bh)$.

\begin{algorithm}[htbp] 
\caption{Differential Evolution for Thresholds} 
\label{alg:de-thresholds}  
\begin{algorithmic}[1] 
\Input Number of thresholds $K$; histograms $\bh$ on the domain $\Omega=[a,b]$; loss $L(\cdot):=L(\cdot;\bh)$ population size $P$; crossover rate $C_r$; max iterations $G_{\max}$; tolerance $\varepsilon > 0$.
\Output Optimal thresholds $\bt_\star^{(g)}$
\State Feasible set $\mathcal{C}=\{\bt\in\Omega^{K}: t_1\le \ldots \le t_{K}\}$. Define $g(\bt)=(t_2-t_1,\ldots,t_K-t_{K-1})^\top$ and the nonnegative violation vector $g_+(\bt)=\max\{0,g(\bt)\}$ (componentwise).
\State \textit{Initialize:} Sample $\bt_p^{(0)}$, $p=1,\ldots,P$, i.i.d. on $\Omega^K$; set $\Pcal^{(0)} \gets \{\bt_1^{(0)},\ldots,\bt_P^{(0)}\}$. Let $\bt_\star^{(0)}$ be the best feasible member if any ($g_+(\bt)=\mathbf{0}$) by $L(\bt)$; if none is feasible, take $\arg\min_{\bt\in\Pcal^{(0)}} \|g_+(\bt)\|_1$.
\For{$g=1,\dots,G_{\max}$} \For{$p=1,\dots,P$}
    \State \textit{Mutation}: Draw distinct $r_1$, $r_2$ from $\{1, \dots, p-1, p+1, \dots, P\}$; sample  $F_m\sim \mathcal{U
}(0.5, 1)$; set $$\bv \leftarrow \bt_\star^{(g-1)} + F_m\big(\bt_{r_1}^{(g-1)}-\bt_{r_2}^{(g-1)}\big)$$
    \State \textit{Crossover}: Draw $j_{\mathrm{rand}}\in\{1,\ldots,K\}$ and $z_k\sim\mathrm{Unif}(0,1)$, $k=1,\ldots,K$, i.i.d.; set 
          \[
            u_k \leftarrow \begin{cases}
              v_k, & z_k \le C_r \text{ or } k=j_{\mathrm{rand}},\\
              (\bt_p^{(g-1)})_k, & \text{otherwise.}
            \end{cases} 
          \]
    \State \textit{Selection:} with $\bu$ the trial and $\bx\!=\!\bt_p^{(g-1)}$, 
          \[
          \bt_p^{(g)}\leftarrow
          \begin{cases}
            \bu, & g_+(\bu)=\mathbf{0},\ g_+(\bx)=\mathbf{0},\ L(\bu)\le L(\bx),\\
            \bx, & g_+(\bu)=\mathbf{0},\ g_+(\bx)=\mathbf{0},\ L(\bu)> L(\bx),\\
            \bu, & g_+(\bu)=\mathbf{0},\ g_+(\bx)\neq\mathbf{0},\\
            \bu, & g_+(\bu)\preceq g(\bx)\ \text{and}\ g_+(\bu)\neq g_+(\bx),\\
            \bx, & \text{otherwise,}
          \end{cases}
          \]
\EndFor 
\State $\Pcal^{(g)} = \{\bt_1^{(g)},\ldots, \bt_P^{(g)}\}$; \ $\bt_\star^{(g)}\gets \argmin_{\bt\in\Pcal^{(g)}} L(\bt)$;
\State \textbf{if } {$\mathrm{std}\bigl\{L\bigl(\bt_p^{(g)}\bigr)\bigr\}_{p=1}^P \le \varepsilon\cdot\bigl|\frac1P\sum_p L\bigl(\bt_p^{(g)}\bigr)\bigr|$} \textbf{ then break}
\EndFor 
\end{algorithmic} 
\end{algorithm}

\end{document}